\documentclass[fleqn,usenatbib,onecolumn]{mnras}

\usepackage{newtxtext,newtxmath}

\usepackage[T1]{fontenc}

\DeclareRobustCommand{\VAN}[3]{#2}
\let\VANthebibliography\thebibliography
\def\thebibliography{\DeclareRobustCommand{\VAN}[3]{##3}\VANthebibliography}

\usepackage{graphicx}	
\usepackage{amsmath}	
\usepackage{color}
\usepackage[version=3]{mhchem}

\title[Resolving Dust budget crisis at $z\sim 8$]{Resolving the Dust Budget Crisis at $z \sim 8$ with Optically Thick, High-Density Molecular Clumps: MACS0416\_Y1}

\author[R. R. Kano et al.]{
Ryusei R.\ Kano,$^{1, 2}$\thanks{E-mail: ryuseikano@nagoya-u.jp (RRK)}
Tsutomu T.\ Takeuchi, $^{1,3}$
Erina R.\ Kawamoto,$^{1}$
Ryosuke S.\ Asano,$^{1}$
\newauthor
Masato \ Hagimoto,$^{1}$
and Yoichi \ Tamura$^{1}$
\\
$^{1}$Division of Particle and Astrophysical Science, Nagoya University, Furo-cho, Chikusa-ku, Nagoya, 464–8602, Japan\\
$^{2}$The Institute for Astronomy (IfA), School of Physics and Astronomy, the University of Edinburgh, Royal Observatory, Edinburgh, EH9 3HJ, UK\\
$^{3}$The Research Center for Statistical Machine Learning, The Institute of Statistical Mathematics, 10-3 Midori-cho, Tachikawa, Tokyo 190–8562, Japan
}

\date{Accepted XXX. Received YYY; in original form ZZZ}

\pubyear{2026}

\begin{document}
\label{firstpage}
\pagerange{\pageref{firstpage}--\pageref{lastpage}}
\maketitle

\begin{abstract}
Dust plays a crucial role in galaxy evolution by shaping the spectral energy distribution (SED) and star formation history. However, standard models often underestimate the infrared luminosity of high-redshift galaxies ($z \sim 8$), leading to the so-called dust budget crisis. In this work, we modify the theoretical framework by focusing on compact star-forming clumps in the interstellar medium. Motivated by the observed compactness of high-z galaxies, we treat the cold neutral medium density as a free parameter. Our analysis reveals that the ISM must reach extreme densities ($n_{\text{H,CNM}} \sim 7.5 \times 10^3 \, \mathrm{cm}^{-3}$). This enhances UV photon trapping, accelerates dust processing in dense gas, and reduces dust destruction by supernova shocks. Our model successfully reproduces the observed UV-to-FIR SED of MACS0416\_Y1 ($z = 8.312$). A grain-size-resolved treatment further shows that the warm IR emission is dominated by intermediate-size grains ($a = 0.01$--$0.1\,\mu$m), which contribute about 89\% of the luminosity near the SED peak and in the ALMA Band~9 continuum. These grains are nearly in thermal equilibrium at characteristic temperatures of $\sim 70$\,K, while the largest grains remain cooler and the smallest grains exhibit a high-temperature tail with low probability. We conclude that extreme ISM densities can alleviate the dust budget crisis by promoting efficient UV photon trapping and rapid dust evolution, thereby increasing dust mass and producing a multi-temperature grain population.
\end{abstract}




\section{Introduction}
\label{sec:introduction}

Galaxies are complex systems composed of gas, stars, dust, and dark matter, all of which emit light across various wavelengths. Analyzing their spectral energy distribution (SED), which captures the energy emitted across the electromagnetic spectrum, is essential for extracting physical information about galaxies. The SED provides crucial insights into a galaxy's star formation history and the properties of its interstellar dust---key factors governing galactic evolution. Since dust grains are formed from heavy elements produced by stars, understanding dust evolution inherently requires tracing the chemical enrichment of the galaxy. Therefore, to accurately interpret these SEDs, a theoretical framework consistently coupling chemical and dust evolution is indispensable.

Over the past decades, numerous studies have aimed to theoretically replicate the SEDs of galaxies. Early comprehensive models \citep[e.g.,][]{silva1998modeling, granato2000infrared, popescu2000modelling, da2008simple} successfully reproduced the observed SEDs of nearby spiral and starburst galaxies by incorporating radiative transfer and dust reprocessing. Building on these foundations, recent advancements have integrated more sophisticated dust physics. For instance, \citet{nishida2022new} successfully modeled the SEDs of nearby galaxies ($z \sim 0$) from ultraviolet (UV) to infrared (IR) wavelengths by employing the dust evolution model developed by \citet{asano2013dust, asano2013determines} (hereafter the Asano model) and \citet{nozawa2015evolution}. The Asano model provides a comprehensive representation of dust life-cycles, including grain formation in stellar winds and supernovae, growth in the interstellar medium (ISM), and destruction by shock waves \citep{Hirashita2011EffectsOG, hou2016dust}. These models, calibrated to the local universe, have proven effective in reproducing the properties of nearby galaxies. Their applicability to the much more extreme environments of high-redshift galaxies, however, remains uncertain.

When applied to high-redshift galaxies ($z > 6$), these models calibrated to the local universe reveal significant discrepancies. The longstanding ``dust budget crisis,'' initially identified by early studies \citep[e.g.,][]{michalowski2010dust, rowlands2014dust}, has been fundamentally reframed by recent ALMA and JWST observations. While ALMA surveys detected substantial dust masses at $z \simeq 7$ requiring rapid enrichment \citep{sommovigo2022alma, algera2024accurate}, JWST has revealed a contrasting population of dust-poor, UV-bright systems at $z > 10$ \citep{ferrara2025blue}. Reconciling this diverse early universe requires models to account for lower supernova dust survival rates due to shock destruction \citep{vasiliev2024dust}, rapid dust ejection mechanisms \citep{tsuna2023photon}, and consequently, a heavy reliance on efficient ISM grain growth \citep{mauerhofer2023dust}. Even with these advancements, state-of-the-art simulations typically yield dust masses $\sim 1$\,dex lower than standard observational estimates \citep{choban2025dusty}. One possible reason is that these frameworks generally adopt ISM conditions calibrated to the local universe, including a characteristic cold neutral medium (CNM) density of $n_{\mathrm{H,CNM}} \sim 30\,\mathrm{cm}^{-3}$, which may severely underestimate the densities of compact, high-pressure star-forming regions in high-redshift galaxies. To resolve this apparent discrepancy, both theoretical and observational studies increasingly suggest that high-redshift galaxies possess warmer dust temperatures ($T_{\text{dust}} \gtrsim 40$--$80$\,K) compared to local galaxies, which effectively reduces the observationally inferred dust masses \citep{sommovigo2021dust, sommovigo2022alma}.

An illustrative example of this tension in temperature-based interpretations is the galaxy MACS0416\_Y1 at $z = 8.31$ \citep{tamura2019detections, Bakx2020}. Early submillimetre observations, such as the ALMA Band 7 detection, implied a moderate dust mass under the standard assumptions of optically thin emission and a dust temperature of $T_{\mathrm{dust}} = 50\,\mathrm{K}$ \citep{tamura2019detections}. However, subsequent shorter-wavelength constraints, including an ALMA Band 9 detection, indicated a substantially higher total infrared luminosity \citep{Bakx2025}. In the optically thin modified-blackbody framework widely adopted in high-redshift studies \citep[e.g.,][]{blain2002}, the infrared luminosity scales as $L_{\mathrm{IR}} \propto M_{\mathrm{dust}} T_{\mathrm{dust}}^{4+\beta}$ \citep{Hildebrand1983}. Within this framework, reproducing the intense short-wavelength emission requires a very high dust temperature, $T_{\mathrm{dust}} \sim 91\,\mathrm{K}$, in order to keep the inferred dust mass within plausible limits \citep{Bakx2025}, leading to the interpretation of MACS0416\_Y1 as a warm ultra-luminous infrared galaxy (ULIRG).

However, this derivation relies heavily on the optically thin, single-temperature assumption. If the dust geometry is compact and optically thick, and if the dust size distribution is dynamically evolving, this simple proportionality breaks down.

In this work, we propose a physically motivated solution to the dust budget crisis. Instead of solely introducing ad hoc dust production mechanisms, we focus on the density evolution of the ISM. Cosmological scaling relations imply that the virial density of dark matter halos evolves as $(1+z)^3$, and observations confirm that high-$z$ galaxies are extremely compact on galactic scales \citep[e.g.,][]{Shibuya2015}. Crucially, theoretical and observational constraints on the interstellar medium demonstrate that the gas surface density within star-forming regions systematically increases with redshift \citep{sommovigo2021dust}. Driven by these combined macroscopic and ISM-scale conditions, it naturally follows that individual star-forming clumps and molecular clouds in these early epochs are significantly denser than their local counterparts.

In this picture, a high-density environment is expected to drive three related effects, synergistic effects simultaneously: 
(1) it naturally results in a compact, optically thick geometry, which maximizes the efficiency of converting stellar UV radiation into IR emission; 
(2) as we will demonstrate, it drastically compresses the timescale of interstellar processing. Accelerated metal accretion, shattering, and coagulation rapidly reshape the dust population within an exceptionally short timescale ($<80$\,Myr), driving a ``compressed evolution''; and 
(3) it significantly suppresses the efficiency of dust destruction by supernova shocks, as rapid radiative cooling shortens the adiabatic phase where sputtering is most effective \citep[e.g.,][]{Draine1979, McKee1989}.

Our primary objective is to extend the theoretical framework of \citet{nishida2022new} to high-density ISM conditions and apply it to MACS0416\_Y1. Specifically, by treating the ISM density as a free parameter, we investigate whether the observed IR emission can be reproduced. We find that the data statistically prefer substantially higher hydrogen number densities ($n_{\text{H,CNM}} \sim 7.5 \times 10^3 \, \mathrm{cm}^{-3}$, corresponding to a model parameter $n_{0,\text{CNM}} = 2.34 \times 10^5 \, \mathrm{cm}^{-3}$) compared to their local counterparts. We also examine how a size-resolved treatment of dust evolution modifies the resulting grain population and IR emission, and whether the strong ALMA Band 9 emission can be explained without invoking an extremely high single dust temperature. In this way, we aim to provide a physically consistent picture of dust evolution in the dense environments of the early universe.

\section{METHODS: CONSTRUCTION OF GALAXY SED}
\label{sec:methods}

\begin{figure}
\centering
\includegraphics[width=0.48\columnwidth]{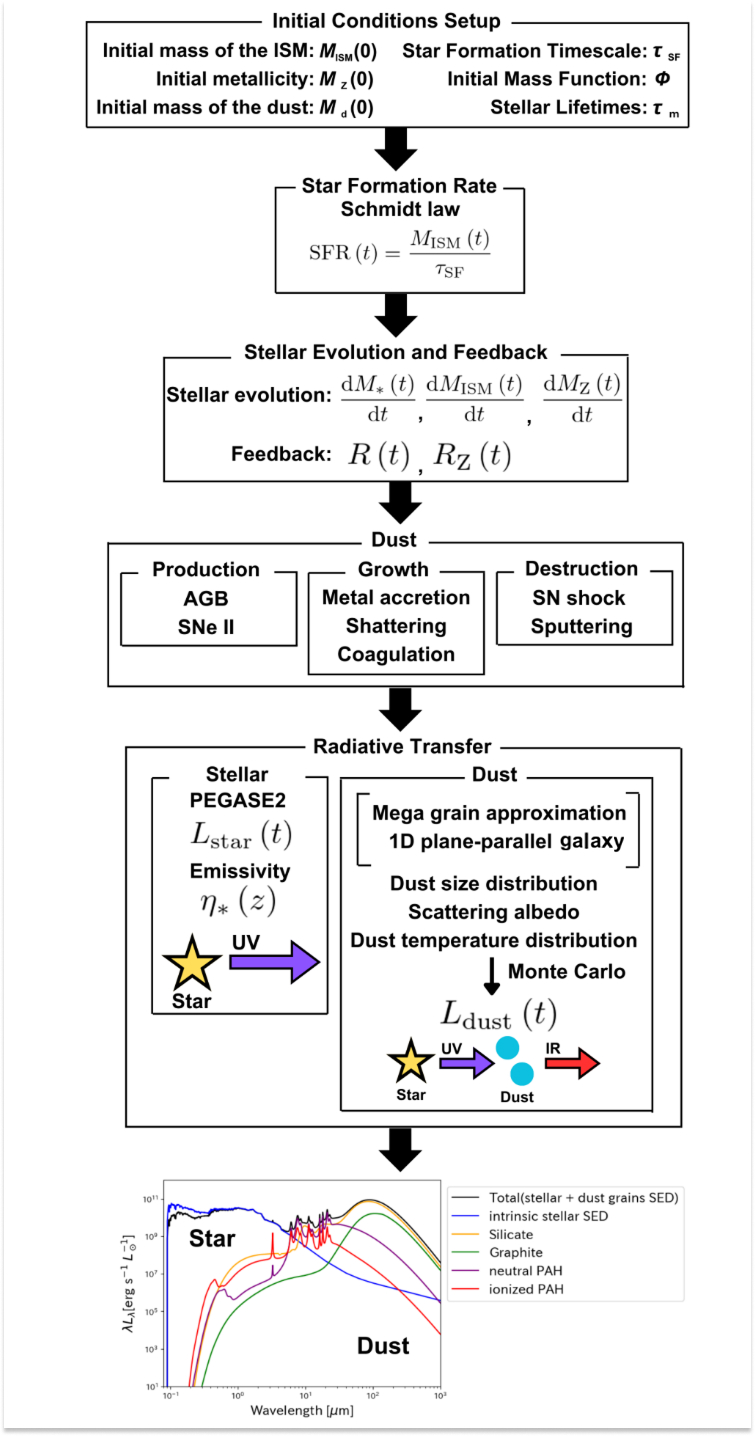}
\caption{The flowchart of construction of galaxy SED.  In the construction of a galaxy's SED, the process begins with the establishment of initial conditions. These include the initial mass of the ISM, $M_{\mathrm{ISM}}(0)$; the initial metallicity, $M_{\mathrm{Z}}(0)$; the initial mass of dust, $M_{\mathrm{d}}(0)$; the star formation timescale, $\tau_{\mathrm{SF}}$; the initial mass function, $\phi$; and the stellar lifetime, $\tau_{\mathrm{m}}$. Utilizing the Schmidt law, the SFR is then calculated. Following the establishment of these parameters, the temporal evolution of the stellar mass, $\mathrm{d}M_{*}(t)/\mathrm{d}t$; the ISM mass, $\mathrm{d}M_{\mathrm{ISM}}(t)/\mathrm{d}t$; and the metal mass, $\mathrm{d}M_{\mathrm{Z}}/(t)\mathrm{d}t$, are computed. Additionally, the return of gas to the ISM, $R(t)$, and the ejection of metals, $R_{\mathrm{Z}}(t)$, are also determined. The next phase considers the evolution of dust, which includes processes such as dust production, destruction, and growth. With these elements defined, the radiative transfer calculations for both stellar and dust components are performed. Ultimately, these steps culminate in the generation of the galaxy's SED.
}
\label{fig:Initial-Conditions-Setup}
\end{figure}

In this section, we describe our framework for reproducing the SEDs of galaxies. Our dust evolution model builds upon the formalism established by \citet{asano2013dust, asano2013determines, asano2014evolution}, while the SED generation methodology follows \citet{nishida2022new}. In Section \ref{subsec:GalaxyEvolution}, we introduce the governing equations of galaxy evolution. Section \ref{subsec:DustEvolution} outlines the fundamental properties of dust and describes the evolution equations incorporated into our model. Finally, in Section \ref{subsec:ConstructSED}, we detail the SED calculation method, including the specific assumptions made for high-redshift environments. Figure \ref{fig:Initial-Conditions-Setup} presents a flowchart of the overall procedure for constructing the galaxy SEDs.

\subsection{Equations of galaxy evolution}
\label{subsec:GalaxyEvolution}

The baryonic content of galaxies consists of stars, gas, and dust grains. We model the evolution of these components by tracking four key parameters: stellar mass $M_*$, ISM mass $M_{\mathrm{ISM}}$, metal mass $M_\mathrm{Z}$, and dust mass $M_\mathrm{d}$. Following the formulation of previous studies \citep[e.g.,][]{lisenfeld1998dust, hirashita1999dust, asano2013dust}, the time evolution of these quantities is described by the following set of equations:

\begin{align}
    \frac{\mathrm{d}M_{*} (t)}{\mathrm{d} t} &= \mathrm{SFR}(t) - R(t), \label{eq:Mstar} \\
    \frac{\mathrm{d} M_{\mathrm{ISM}} (t)}{\mathrm{d} t} &= - \mathrm{SFR}(t) + R(t) + \frac{\mathrm{d} M_{\mathrm{infall}} (t)}{\mathrm{d} t},\\
    \frac{\mathrm{d} M_{\mathrm{Z}} (t)}{\mathrm{d} t} &= -Z(t) \mathrm{SFR}(t) + R_{\mathrm{Z}}(t) + Y_{\mathrm{Z}}(t),\\
    \frac{\mathrm{d} M_{\mathrm{d}} (t)}{\mathrm{d} t} &=-D(t) \mathrm{SFR}(t) + Y_{\mathrm{d}}(t) - \left( \frac{\mathrm{d} M_{\mathrm{d}} (t)}{\mathrm{d} t} \right)_{\mathrm{SN}} + \left( \frac{\mathrm{d} M_{\mathrm{d}} (t)}{\mathrm{d} t} \right)_{\mathrm{acc}},
    \label{eq:simpleDustEvolution}
\end{align}
where $\mathrm{SFR}(t)$ is the star formation rate. $R(t)$ and $R_{\mathrm{Z}}(t)$ represent the rates at which gas and metals are returned to the ISM from stars at the end of their lives, respectively. $\mathrm{d}M_{\mathrm{infall}} / \mathrm{d} t$ denotes the gas infall rate from the intergalactic medium. The metallicity and dust-to-gas ratio are defined as $Z(t) \equiv M_{\mathrm{Z}}/M_{\mathrm{ISM}}$ and $D(t) \equiv M_{\mathrm{d}}/M_{\mathrm{ISM}}$, respectively. $Y_{\mathrm{Z}}(t)$ and $Y_{\mathrm{d}}(t)$ represent the cumulative mass of metals and dust newly synthesized and ejected by stars per unit time. The terms $(\mathrm{d} M_{\mathrm{d}} /\mathrm{d} t)_{\mathrm{SN}}$ and $(\mathrm{d} M_{\mathrm{d}} /\mathrm{d} t)_{\mathrm{acc}}$ denote the rates of change in dust mass due to supernova (SN) shock destruction and grain growth via accretion in the ISM, respectively. 
Notably, $(\mathrm{d} M_{\mathrm{d}} / \mathrm{d} t)_{\mathrm{acc}}$ is calculated based on the continuity equation in the grain size space, as detailed in Section \ref{subsubsec:EquationDustEvolution}.

This section details the terms in Equations \eqref{eq:Mstar} to \eqref{eq:simpleDustEvolution}. For the SFR, we adopt the Schmidt law \citep{schmidt1959rate}, $\mathrm{SFR}\propto M^{n}_{\mathrm{ISM}}$, with an index of $n=1$ for simplicity:
\begin{equation}
    \mathrm{SFR}(t) = \frac{M_{\mathrm{ISM}}(t)}{\tau _{\mathrm{SF}}},
    \label{eq:schmidtLaw}
\end{equation}
where $\tau _{\mathrm{SF}}$ represents the star formation timescale. The terms $R(t)$, $R_{\mathrm{Z}}(t)$, $Y_{\mathrm{Z}}(t)$, and $Y_{\mathrm{d}}(t)$ are calculated as follows:
\begin{align}
    R(t) &=\int _{m_{\min }(t) }^{100\mathrm{M}_{\odot }}\left[ m-\omega \left( m,Z(t-\tau_m) \right) \right] \phi(m) \mathrm{SFR}(t-\tau_m) \mathrm{d}m,\\
    R_{\mathrm{Z}}(t) &=\int _{m_{\min }(t) }^{100\mathrm{M}_{\odot }}\left[ m-\omega \left( m,Z(t-\tau_m) \right) \right] \phi(m)
    \mathrm{SFR}(t-\tau_m) Z(t-\tau_m) \mathrm{d}m, \\
    Y_{\mathrm{Z}}(t) &=\int _{m_{\min }(t) }^{100 \mathrm{M}_{\odot }}m_{\mathrm{Z}}\left( m,Z(t-\tau_m) \right) \phi(m) \mathrm{SFR}(t- \tau_m) \mathrm{d}m, \\
    Y_{\mathrm{d}}(t) &=\int _{m_{\min }(t) }^{100 \mathrm{M}_{\odot }}m_{\mathrm{d}}\left( m,Z(t-\tau_m) \right) \phi(m) \mathrm{SFR}(t- \tau_m) \mathrm{d}m,
\end{align}
where $m_{\mathrm{min}}(t)$ is the minimum stellar mass contributing to the feedback at time $t$, and $\phi(m)$ is the initial mass function (IMF). 
$\omega(m, Z)$, $m_{\mathrm{Z}}(m, Z)$, and $m_{\mathrm{d}}(m, Z)$ represent the remnant mass, the newly synthesized metal mass, and the newly formed dust mass, respectively, for a star of initial mass $m$ and metallicity $Z$.

The stellar lifetime $\tau_{m}$ for a star of mass $m$ is adopted from \citet{raiteri1996simulations}:
\begin{align}
\log \tau _{m} &= a_{0}(Z) +a_{1}(Z) \log m + a_{2}(Z) (\log m)^{2},
\end{align}
with the metallicity-dependent coefficients:
\begin{align}
    a_0(Z) &= 10.13 + 0.07547 \log Z - 0.008084 (\log Z)^2, \\
    a_1(Z) &= -4.424 -0.7939 \log Z - 0.1187 (\log Z)^2, \\
    a_2(Z) &= 1.262 +0.3385 \log Z + 0.05417 (\log Z)^2.
\end{align}
These coefficients are derived from the stellar evolutionary tracks of the Padova group \citep{alongi1993evolutionary, bressan1993evolutionary, bertelli1994theoretical} for a mass range of $0.6\text{--}120\,\mathrm{M}_{\odot}$ and metallicity range of $0.0004\text{--}0.05 \mathrm{Z_\odot}$. 

In this study, we utilize the Chabrier IMF \citep{chabrier2003galactic}. Compared to the Salpeter IMF \citep{salpeter1955luminosity}, the Chabrier IMF is characterized by a turnover at lower masses, resulting in a smaller fraction of low-mass stars per unit mass formed. The IMF is formulated as:
\begin{equation}
\label{eq:ChabrierIMF}
  \phi(m) \propto
  \begin{cases}
    \frac{1}{m} \exp \left[- \frac{(\log m - \log m_{\mathrm{c}})^2}{2 \sigma ^2} \right] & (m \le 1 \mathrm{M}_{\odot}) \\
    m^{-2.3} & (m > 1 \mathrm{M}_{\odot}),
  \end{cases}
\end{equation}
where $m_{\mathrm{c}} = 0.08\,\mathrm{M}_{\odot}$ is the characteristic mass and $\sigma = 0.69$ is the width of the distribution. The IMF is normalized such that
\begin{equation}
\label{eq:ChabrierIMFNormalized}
    \int ^{100 \mathrm{M}_{\odot}} _{0.1 \mathrm{M_\odot}} m \phi(m) \mathrm{d} m = 1 \mathrm{M}_\odot.
\end{equation}

For the infall model, we adopt the following equation \citep{inoue2011origin}:

\begin{equation}\label{eq:infall_model}
    \frac{\mathrm{d}M_{\mathrm{infall}}}{\mathrm{d}t} = \frac{M_{\mathrm{infall}}}{\tau_{\mathrm{infall}}}\exp\left( - \frac{t}{\tau_{\mathrm{infall}}} \right),
\end{equation}
where  $M_{\mathrm{infall}}$ is the total mass that flows into the galaxy by infall as $t \rightarrow \infty$. The initial mass of a galaxy is assumed to be negligible.

\subsection{Dust evolution model}
\label{subsec:DustEvolution}

Interstellar dust plays a pivotal role in galaxy evolution, regulating the thermodynamic state of the gas and shaping the SED by absorbing UV photons and re-emitting energy in the IR. Additionally, dust grains catalyze the formation of molecular hydrogen ($\mathrm{H}_2$), thereby influencing the star formation efficiency \citep[e.g.,][]{calzetti2001dust, hirashita2002dust}. In this study, we employ the dust evolution framework developed by \citet{asano2013dust, asano2013determines, asano2014evolution} and \citet{nozawa2015evolution}. While the original model tracks ten dust species (\ce{C}, \ce{Si}, \ce{SiO2}, \ce{Fe}, \ce{FeS}, \ce{Al2O3}, \ce{MgO}, \ce{MgSiO3}, \ce{Mg2SiO4}, and \ce{Fe3O4}), we categorize them into two principal families for simplicity: astronomical silicate and carbonaceous grains. This simplification facilitates the computation of grain--grain processing such as coagulation and shattering. We compute the grain size distribution and collisional processing (coagulation and shattering)
using two grain families, carbonaceous and silicate. Iron-bearing dust is not treated as an
independent collisional family; instead, we associate it with the silicate family and assume
the same collisional and kinematic parameters (e.g. relative velocities and collision rates)
as those used for silicate grains. For bookkeeping, however, we separately track the dust mass
in the iron component while it shares the size distribution and collisional evolution of the
silicate family. Throughout this paper, unless otherwise specified, the calculated and observed SEDs span the wavelength range from the rest-frame UV to the FIR/submillimeter ($\lambda_{\rm rest} \sim 0.1 - 1000\mu\mathrm{m}$). Section \ref{subsubsec:DustScenario} outlines our dust evolution scenario, and Section \ref{subsubsec:EquationDustEvolution} details the governing equations.

\subsubsection{Dust evolution scenario}
\label{subsubsec:DustScenario}
Our model incorporates key dust processing mechanisms: dust production by asymptotic giant branch (AGB) stars and Type II supernovae (SNe II), destruction by SN shocks, grain growth via metal accretion in the CNM and molecular clouds (MC), and grain--grain collisions (shattering and coagulation) in the warm neutral medium (WNM), CNM, and MC phases. Most model parameters are adopted from \citet{nishida2022new}, as summarized below.

AGB stars, representing the final evolutionary stage of low- to intermediate-mass stars ($< 8\,\mathrm{M}_\odot$), enrich the ISM with heavy elements and dust grains. We model the dust size distribution as a lognormal function with a peak at $0.1\,\mathrm{\mu m}$ and a standard deviation of $\sigma = 0.47$, consistent with \citet{yasuda2012formation}. Dust yields are calculated based on \citet{zhukovska2008evolution} for stellar masses of $1\text{--}8\,\mathrm{M}_\odot$ and metallicities of $0.05\text{--}0.4\,\mathrm{Z_\odot}$, extrapolated to cover the full parameter space. It is worth noting that at the high redshifts of interest ($z \gtrsim 8$), the contribution of AGB stars is expected to be sub-dominant compared to SNe II, as the evolutionary timescales of low-mass stars are comparable to or longer than the age of the Universe at these epochs.

For SNe II (progenitor masses $13\text{--}30\,\mathrm{M}_{\odot}$), we adopt the dust size distributions for unmixed progenitors from \citet{nozawa2007evolution}. Following the explosion, the reverse shock traversing the ejecta destroys a fraction of the newly formed dust via sputtering \citep[e.g.,][]{nozawa2006dust, bianchi2007dust}. We account for both thermal sputtering (dependent on gas temperature and density) and non-thermal sputtering (driven by the relative velocity between gas and grains).

In the dense ISM (CNM and MC phases), dust mass increases via metal accretion (grain growth) \citep[e.g.,][]{dwek1980evolution, Hirashita2011EffectsOG}. As noted by \citet{asano2013dust}, this process becomes efficient only after the ISM is sufficiently enriched with metals (typically after $\sim 1$\,Gyr in local galaxies). 
For our baseline model representing local environments, we follow \citet{nishida2022new} and assume grain growth occurs in the CNM ($T_{\mathrm{gas}} = 100\,\mathrm{K}$, $n_{\mathrm{H}} = 30\,\mathrm{cm}^{-3}$) and MC ($T_{\mathrm{gas}} = 25\,\mathrm{K}$, $n_{\mathrm{H}} = 300\,\mathrm{cm}^{-3}$), with a total mass fraction of 0.5 for these phases \citep{nozawa2015evolution}. 
However, as we discuss in Section \ref{subsec:resolution_high_density}, these density parameters are treated as variables to account for the compact and high-pressure conditions of high-redshift galaxies. Grains are assumed to grow as spheres, and we consider their geometric cross-sections.

Turbulence in the ISM induces relative velocities between grains \citep{yan2004dust}, leading to collisions. These are categorized into shattering (high-velocity collisions creating smaller fragments) and coagulation (low-velocity collisions leading to growth). While these processes conserve total dust mass, they significantly alter the grain size distribution. We adopt the grain velocity models of \citet{yan2004dust}, considering collisions only between grains of the same species.

\subsubsection{Equation of dust evolution}
\label{subsubsec:EquationDustEvolution}
We solve the time evolution of the dust mass density $\rho_{\mathrm{d,X}}(a,t)$ for a grain species $X$ within a size bin $[a, a+\mathrm{d}a]$. The evolution equation is given by:
\begin{align}
    \frac{\mathrm{d} \rho_{\mathrm{d,X}} (a,t)}{\mathrm{d} t} = \left[ \frac{\mathrm{d} \rho_{\mathrm{d,X}} (a,t)}{\mathrm{d} t} \right]_{\mathrm{inj}} 
    + \left[ \frac{\mathrm{d} \rho_{\mathrm{d,X}} (a,t)}{\mathrm{d} t} \right]_{\mathrm{dest}}
    + \left[ \frac{\mathrm{d} \rho_{\mathrm{d,X}} (a,t)}{\mathrm{d} t} \right]_{\mathrm{growth}} 
    + \left[ \frac{\mathrm{d} \rho_{\mathrm{d,X}} (a,t)}{\mathrm{d} t} \right]_{\mathrm{grain-grain}}
\end{align}
where the terms on the right-hand side represent the rates of change due to stellar injection (wind/SN), shock destruction, grain growth, and grain--grain collisions, respectively.

The injection term includes dust production and astration:
\begin{equation}
    \left[ \frac{\mathrm{d} \rho_{\mathrm{d,X}}(a,t)}{\mathrm{d} t} \right]_{\mathrm{inj}} = - \frac{\rho_\mathrm{d,X}(a,t)}{M_{\mathrm{ISM}}(t)} \mathrm{SFR}(t) + Y_{\mathrm{d,X}}(a,t),
\end{equation}
where the ejection rate $Y_{\mathrm{d,X}}(a,t)$ is
\begin{align}
    Y_{\mathrm{d,X}}(a,t) =& \int_{m_{\mathrm{min}}(t)}^{100 \mathrm{M}_{\odot}} m_{\mathrm{d}} \left[ m, Z(t- \tau_{m}), a \right] \phi(m) \mathrm{SFR}(t-\tau_m) \,\mathrm{d}m.
\end{align}

The shock destruction term is given by
\begin{align}
    \left[ \frac{\mathrm{d} \rho_{\mathrm{d,X}}(a,t)}{\mathrm{d} t} \right]_{\mathrm{dest}} = - \frac{M_{\mathrm{swept}}}{M_{\mathrm{ISM}}(t)}\gamma_{\mathrm{SN}}(t)  \left[\rho_{\mathrm{d,X}}(a,t) - m(a) \int_{0}^{\infty} \eta_{\mathrm{X}}(a,a^\prime) f_{\mathrm{X}}(a^\prime,t) \, \mathrm{d}a^\prime \right],
\end{align}
where $M_{\rm swept}$ is the ISM mass swept by a SN shock. $M_{\rm swept}$ depends on the density and metallicity of the ISM \citep{nozawa2006dust, yamasawa2011role}. When the ISM density is high, $M_{\rm swept}$ is small because there are many particles that slow down the SN shock. When the metallicity is high, efficient line cooling with metal results in a faster shock deceleration and smaller $M_{\rm swept}$. We use the following formulae fitted by \citet{yamasawa2011role},
\begin{equation}
    M_{\rm swept} = 1535 n_{\rm SN}^{-0.202} \left[ (Z/Z_{\odot}) + 0.039 \right]^{-0.289} \mathrm{M}_{\odot},
    \label{eq:m_swept}
\end{equation}
where $n_{\rm SN}$ is the ISM density surrounding SNe.

The grain growth via metal accretion and grain--grain collisions (shattering and coagulation) are also included in our model. For the detailed mathematical formulations of these processes, we strictly follow the methods detailed in \citet{asano2013dust} and \citet{nishida2022new}.
\subsection{Construction of SED}
\label{subsec:ConstructSED}
In galaxies, the spectral energy source is divided into stellar, gas and dust. First, radiation from stars is calculated by simple stellar population (SSP) method. Second, the dust size distribution, dust properties and the dust attenuation is treated by chemical evolution model mentioned in Section \ref{subsec:DustEvolution}. Third, the dust emission is calculated by stochastic heating method. Finally, the galaxy SED is calculated by combining the stellar SED, the attenuation curve, and the dust emission.

\subsubsection{Stellar SED model}
In our SED model, we use the same method as $\text{P}\acute{\mathrm{E}}\text{GASE}$ \citep{fioc1999pegase}, SSP method for stellar SED. By considering SSP, the monochromatic luminosity per unit wavelength is represented as
\begin{equation}\label{eq:SSP}
    L^{\mathrm{SSP}}_{\lambda} \left( t, Z \right) = \int_{m_{\mathrm{min}}}^{m_{\mathrm{max}}\left( t \right)} L_{\lambda}^{\mathrm{star}}\left( T_{\mathrm{eff}} \left( t,m \right), \log g \left( t,m \right), Z \right) \phi \left( m \right) d \left( \ln m \right),
\end{equation}
where $t$ is the age, $Z$ is the metallicity, $\lambda$ is the wavelength, $L_{\lambda}^{\mathrm{star}}$ is the monochromatic luminosity of a star with mass in the interval $[m, m+ \mathrm{d}m]$, effective temperature $T_{\mathrm{eff}}$, and surface gravity $g$.  In this paper, we set 
 $m_{\mathrm{min}}=0.1 \mathrm{M}_{\odot}$, and $m_{\mathrm{max}}\left( t \right)=100 \mathrm{M}_\odot$. 

The total stellar luminosity spectrum is computed by integrating the SSP contributions over the star formation history:
 \begin{equation}\label{eq:total_monochromatic_luminosity}
 L_{\mathrm{\lambda}}\left( t \right) = \int_{t^\prime=0}^{t^\prime=t} \int_{Z=0}^{Z=Z_{\mathrm{max}}} \mathrm{SFR}\left( t-t^\prime \right) P \left( t-t^\prime,Z \right) L_{\lambda}^{\mathrm{SSP}} \left( t^\prime, Z \left[ t-t^\prime \right] \right) \,\mathrm{d}t^\prime  \,\mathrm{d}Z, 
 \end{equation}
 where $P \left( t-t^\prime,Z \right)$ is the time-dependent metallicity distribution function.  The SFR and IMF is defined in Section \ref{subsec:GalaxyEvolution}.

\subsubsection{Mega grain approximation}
\label{subsubsec:MGA}
We use mega grain approximation (MGA) by \citet{varosi1999analytical} and \citet{inoue2005attenuation} to calculate radiative transfer for reducing computing costs in a one-dimensional plane parallel galaxy. In MGA, we treat the dusty region as a kpc-scale homogeneous medium called mega-grain. The clumps are presumed to be the central regions of molecular clouds and the birthplaces of young stars. 
\citet{inoue2005attenuation} assumed the size of clumps is determined by the self-gravity of the CNM:
\begin{equation}
    r_{\text{cl}} = \frac{1}{\rho_{\text{cl}}}\sqrt{\frac{15p}{4\pi G}} = \frac{1}{\mu m_{\text{p}} n_{\mathrm{H,CNM}}}\sqrt{\frac{15p}{4\pi G}}.
    \label{eq:clump_radius}
\end{equation}
where $p$ is the thermal pressure of the ISM, $G$ is gravitational constant, $\rho_{\mathrm{cl}}=\mu m_{\mathrm{p}}n_{\mathrm{H,CNM}}$ is the density of clump, $\mu =1.4$ is the mean atomic weight, $m_{\mathrm{p}}= 1.673 \times 10^{-24}\mathrm{g}$ is the proton mass and $n_{\mathrm{H,CNM}}$ is the hydrogen density of CNM. We assume pressure equilibrium between the WNM and CNM phases, where the pressure is related to the CNM density as:
\begin{equation}
    \frac{p}{k_{\mathrm{B}}} = 10^{4.5} \left( \frac{n_{\mathrm{H,CNM}}}{n_{0,\mathrm{CNM}}} \right)^{0.7} \, \mathrm{K \, cm^{-3}},
    \label{eq:pressure_equilibrium}
\end{equation}
where $n_{0,\mathrm{CNM}}$ is a reference density that sets the normalization of the
pressure--density relation. We adopt $n_{0,\mathrm{CNM}}=10^{3}\,\mathrm{cm^{-3}}$
as a fiducial value for local galaxies \citep{nishida2022new}, but treat $n_{0,\mathrm{CNM}}$ as a
free parameter for high-$z$ systems throughout this work (see Section~\ref{subsec:MCMCRESULT}).
This relation implies that higher densities lead to higher pressures, which in turn results in smaller clump radii under the self-gravity assumption.

 We adopt the simplifying assumption that all clumps are spherically shaped with uniform density and radius. Based on this assumption, we calculate the mass absorption and scattering coefficients, as well as the scattering asymmetry parameter for dust grains, denoted as $k_\mathrm{abs}$, $k_\mathrm{scat}$, and $g_\mathrm{d}$, respectively. These parameters are averaged over the dust size distribution as computed by the Asano model:
\begin{align}
    k_{\mathrm{abs}} &= \frac{\int_{a_{\mathrm{min}}}^{a_{\mathrm{max}}} \pi a^2 Q_{\mathrm{abs}}\left(a \right) f \left( a \right) \mathrm{d}a}{\int_{a_{\mathrm{min}}}^{a_{\mathrm{max}}} m_{\mathrm{d}} \left( a \right)
    f \left( a \right) \mathrm{d}a} ,\\
    k_{\mathrm{scat}} &= \frac{\int_{a_{\mathrm{min}}}^{a_{\mathrm{max}}} \pi a^2 Q_{\mathrm{scat}}\left(a \right) f \left( a \right) \mathrm{d}a}{\int_{a_{\mathrm{min}}}^{a_{\mathrm{max}}} m_{\mathrm{d}} \left( a \right)
    f \left( a \right) \mathrm{d}a} ,\\
    g_{\mathrm{d}} &= \frac{\int_{a_{\mathrm{min}}}^{a_{\mathrm{max}}} g \left( a \right) \pi a^2 Q_{\mathrm{scat}}\left(a \right) f \left( a \right) \mathrm{d}a}{\int_{a_{\mathrm{min}}}^{a_{\mathrm{max}}} \pi a^2 Q_{\mathrm{scat}}\left(a \right) f \left( a \right) \mathrm{d}a} ,
\end{align}
where $f(a)$ represents the dust number distribution, while $Q_{\mathrm{abs}}(a)$ and $Q_{\mathrm{scat}}(a)$ denote the absorption and scattering coefficients of grain, respectively. Additionally, $g(a)$ is the scattering asymmetry parameter for a grain. In this model, the values of $Q_{\mathrm{abs}}(a)$, $Q_{\mathrm{scat}}(a)$, and $g(a)$ are computed using Mie theory \citep{bohren2008absorption}. We adopt the optical parameters for silicate and graphite from \citet{draine1984optical} and \citet{laor1993spectroscopic}, respectively, and for polycyclic aromatic hydrocarbons (PAHs) from \citet{li2001infrared}. For the radiative transfer/SED calculations, we do not assign separate optical constants to iron dust;
as in \citet{nishida2022new}, the iron component is included in the silicate family and the its optical constants are used to evaluate extinction coefficient and emission. The mass extinction coefficient and scattering albedo are defined as 
\begin{equation}
    k_{\mathrm{d}} = k_{\mathrm{abs}} + k_{\mathrm{scat}}.
\end{equation} 
The relative optical depth of clump with interclump medium is
\begin{equation}\label{eq:optical_depth_clump}
    \tau_{\mathrm{cl}}= \left( \rho_{\mathrm{cl}}-\rho_{\mathrm{icm}}\right)k_{\mathrm{d}}Dr_{\mathrm{cl}} ,
\end{equation}
where $\rho_{\mathrm{icm}}=\mu m_{\mathrm{p}}n_{\mathrm{H,wnm}}$ is the gas density,
$n_{\mathrm{H,wnm}}$ is hydrogen number density of WNM, $k_{\mathrm{d}}$ is the dust opacity, and $D$ is dust-to-gas mass ratio, calculated by the Asano model. The effective extinction coefficient is expressed as
\begin{equation}
    \kappa _{\mathrm{eff}} = \kappa_{\mathrm{mg}} + \kappa_{\mathrm{icm}},
\end{equation}
where $\kappa_{\mathrm{mg}}$ is the extinction coefficient per unit length of the medium by clump, and $\kappa_{\mathrm{icm}}$ is the extinction coefficient of interclump medium represented as
\begin{align}
    \kappa_{\mathrm{mg}} &= n_{\mathrm{cl}} \pi r_{\mathrm{cl}}^2 P_{\mathrm{int}} \left( \tau _{\mathrm{cl}} \right) = \frac{3 f_{\mathrm{cl}}}{4 r_{\mathrm{cl}}} P_{\mathrm{int}} \left( \tau_{\mathrm{cl}} \right), \\
    \kappa_{\mathrm{icm}} &= k_{\mathrm{d}} D \rho_{\mathrm{icm}},
\end{align}
where $n_{\mathrm{cl}}$ is the number density of clump, $f_{\mathrm{cl}}$ is the clump filling fraction, and $n_{\mathrm{H}}$ is the volume-averaged hydrogen number density of the galaxy,
\begin{equation}
    f_{\mathrm{cl}} = \frac{n_{\mathrm{H}}-n_{\mathrm{H,WNM}}}{n_{\mathrm{H,CNM}}-n_{\mathrm{H,WNM}}}.
\end{equation}
Note that while standard models fix $n_{\mathrm{H,CNM}}$ to this local value, our study will treat the clump density as a variable to represent high-redshift environments (see Section \ref{sec:discussion}).
The scattering albedo of a clump is
\begin{equation}
    \omega _{\mathrm{cl}} = \omega_{\mathrm{d}} P_{\mathrm{esc}} \left( \tau_{\mathrm{cl}}, \omega_{\mathrm{d}} \right),
\end{equation}
where $\omega_{\mathrm{d}} = k_{\mathrm{scat}}/k_{\mathrm{d}}$ is the scattering albedo of normal grain averaged by grain size distribution and
\begin{equation}
\label{eq:photonEscapeProbability}
    P_{\mathrm{esc}} \left( \tau, \omega \right) = \frac{\frac{3}{4 \pi}P_{\mathrm{int} \left( \tau \right)}}{1-\omega \left[ 1- \frac{3}{4 \pi} P_{\mathrm{int}} \left( \tau \right) \right]}
\end{equation}
is the photon escape probability from a sphere clump. $P_{\mathrm{int}} \left( \tau \right)$ is the interaction probability against parallel light by a sphere with optical depth $\tau$, and represented as
\begin{equation}
    P_{\mathrm{int}} \left( \tau \right) = 1- \frac{1}{2 \tau^2} + \left( \frac{1}{\tau} + \frac{1}{2 \tau^2} \right) e^{-2 \tau}.
\end{equation}
The effective scattering albedo is
\begin{equation}
    \omega _{\mathrm{eff}} = \frac{\omega_{\mathrm{cl}} \kappa_{\mathrm{mg}}+\omega_{\mathrm{d}} \kappa_{\mathrm{icm}}}{\kappa_{\mathrm{eff}}}
\end{equation}

An asymmetry parameter of clump $g_{\mathrm{cl}}$, and optical parameter that indicates in which direction light escapes, is given by fitting the Monte Carlo calculation result in \citet{varosi1999analytical}, and represented by the following empirical formula,
\begin{equation}
    g_{\mathrm{cl}} \left( \tau_{\mathrm{cl}}, \omega_{\mathrm{cl}}, g_{\mathrm{d}}  \right) = g_{\mathrm{d}} - C \left( 1- \frac{1+e^{-B/A}}{1+e^{\left( \tau_{\mathrm{cl}} -B \right) / A}} \right),
\end{equation}
where
\begin{align}
    A & \equiv 1.5 + 4 g_{\mathrm{d}} ^3 + 2 \omega _{\mathrm{d}} \sqrt{g_{\mathrm{d}}} \exp{-5g_{\mathrm{d}}}, \\
    B & \equiv 2 - g_{\mathrm{d}} \left( 1-g_{\mathrm{d}} \right) - 2 \omega_{\mathrm{d}} g_{\mathrm{d}}, \\
    C & \equiv \frac{1}{3 - \sqrt{2g_{\mathrm{d}}}-2\omega_{\mathrm{d}} g_{\mathrm{d}} \left( 1-g_{\mathrm{d}} \right)}.
\end{align}
The effective asymmetry parameter is
\begin{equation}
    g_{\mathrm{eff}} = \frac{g_{\mathrm{cl}} \kappa_{\mathrm{mg}}+ g_{\mathrm{d}} \kappa_{\mathrm{icm}}}{\kappa_{\mathrm{eff}}}.
\end{equation}

\subsubsection{Radiative transfer in a one-dimensional plane parallel galaxy}
Using MGA in Section \ref{subsubsec:MGA} and a one-dimensional plane parallel galaxy \citep{inoue2005attenuation}, radiative transfer equation is solved. Below, we describe the assumptions made in solving radiative transfer.
\begin{enumerate}
    \item We assume a one-dimensional, plane-parallel galaxy.
    \item The z-axis is perpendicular to the galactic plane, with $z=0$ defined as the center of the galaxy. 
    We assume axially symmetric boundaries.
    \item The galaxy comprises two disk components, Disk1 and Disk2, which share the same center. Disk1 consists of gas, dust, and young stars with a disk thickness of $2h_{\mathrm{d}}$.
    Disk2 is composed solely of older, spread-out stars with a disk thickness of $h_{\mathrm{s}}$.
    \item 
    In Disk1, the densities of gas, dust, and stars are assumed to be constant.
\end{enumerate}
Through these approximations, the optical depth can be defined, and the radiative transfer equation can be described using the radiative intensity. From this radiative transfer equation, it is possible to determine the effective emissivity of stars embedded in clumps using boundary conditions of radiative intensity and assumed emissivity distribution of stars. Consequently, as the distance $z$ from the galactic plane increases, the stellar emission exponentially decreases. These points will be elaborated in the following sections.

We define optical depth $\tau$, using a constant effective extinction coefficient, $\kappa_{\mathrm{eff}}$, as follows:
\begin{equation}
\mathrm{d}\tau = -\kappa_{\mathrm{eff}} \mathrm{d}z,
\end{equation}
where $\tau = 0$ at $z = h_{\mathrm{d}}$, and $\tau = \kappa_{\mathrm{eff}} h_{\mathrm{d}}$ at $z = 0$. Here, $2h_{\mathrm{d}}$ represents the thickness of Disk 1, while Disk 2, composed solely of exponentially distributed old stars, has a thickness of $4h_{\mathrm{d}}$, twice that of Disk 1. The centers of these two disks are aligned.

Radiative transfer in this configuration is formulated by the following equation:
\begin{equation}
\label{eq:radiativeTransfer}
\mathrm{\mu} \frac{\mathrm{d}I (\tau, \mathrm{\mu})}{\mathrm{d}\tau} = -I (\tau, \mathrm{\mu}) + S (\tau, \mathrm{\mu}),
\end{equation}
where $I(\tau, \mathrm{\mu})$ represents the specific intensity at $\tau$ and $\mathrm{\mu} \equiv \cos \theta$, with $\theta$ being the angle between the ray and the $z$-axis. The source function $S$, is given by:
\begin{equation}
\label{eq:sourceFunction}
S(\tau, \mu) = \frac{\eta_*(\tau)}{\kappa_{\mathrm{eff}}} + \omega_{\mathrm{eff}} \int_{-1}^1 I(\tau, \mathrm{\mu}') \Phi(g_{\mathrm{eff}}, \mathrm{\mu}, \mathrm{\mu}') \mathrm{d\mu}',
\end{equation}
where $\eta_*$ denotes stellar emissivity, and $\Phi$ represents the scattering phase function, with the Henyey-Greenstein phase function being adopted here \citep{henyey1941diffuse}.

The first term on the right-hand side of Equation (\ref{eq:sourceFunction}) accounts for the light emitted from stars that escapes directly from the clump where the star was born. The second term represents the internally scattered light within the disk, considered in the direction $\mathrm{\mu}$.

The boundary conditions for the galaxy at $z=0$ and $z=h_{\mathrm{d}}$ are established as follows:
\begin{align}
I(\tau = \kappa_{\mathrm{eff}} h_{\mathrm{d}}, \mathrm{\mu}) &= I(\tau = \kappa_{\mathrm{eff}} h_{\mathrm{d}}, -\mathrm{\mu}), \\
I(\tau = 0, \mathrm{\mu} < 0) &= - \frac{\int_{h_{\mathrm{d}}}^\infty \eta_* (z) \mathrm{d}z}{\mathrm{\mu}}.
\end{align}

The stellar emissivity is normalized so that:
\begin{equation}
\label{eq:stellarEmissivityNormalized}
\int_{-\infty}^\infty \eta_*(z) , \mathrm{d}z = 1.
\end{equation}

The emissivity from young stars in Disk 1, $\eta_*^{\mathrm{young}}$, equals $1/(2h_{\mathrm{d}})$ for $|z| \le h_{\mathrm{d}}$ and is zero for $|z| > h_{\mathrm{d}}$, normalized as per Equation (\ref{eq:stellarEmissivityNormalized}). The energy emitted by these stars is absorbed by surrounding dust clumps, with the escape probability modeled by Equation (\ref{eq:photonEscapeProbability}). Thus, the young stars' emissivity is represented by:

\begin{equation}
\eta_*^{\mathrm{young}} =
\begin{cases}
P_{\mathrm{esc}} (\tau_{\mathrm{cl}}, \omega_{\mathrm{cl}})/2h_{\mathrm{d}} & (|z| \le h_{\mathrm{d}}) \\
0 & (|z| > h_{\mathrm{d}})
\end{cases}.
\end{equation}

Old stars in Disk 2 are distributed exponentially along the $z$-axis, with their emissivity $\eta_*^{\mathrm{old}}(z)$, modeled as:

\begin{equation}
\eta_*^{\mathrm{old}}(z) = \frac{e^{-|z|/2h_{\mathrm{d}}}}{4h_{\mathrm{d}}}.
\end{equation}

The total stellar emissivity at $z$, $\eta_*(z)$, is a composite of contributions from young and old stars, calculated by:
\begin{equation}
\eta_*(z) = f_{\mathrm{y}}(t) \eta_*^{\mathrm{young}}(z) + (1 - f_{\mathrm{y}}(t)) \eta_*^{\mathrm{old}}(z),
\end{equation}
where $f_{\mathrm{y}}(t)$ is the luminosity fraction emitted by young stars at age $t$, determined by:
\begin{equation}
f_{\mathrm{y}}(t) = \frac{\int_0^{\min[t_{\mathrm{y}}, t]} \int_0^{Z_{\mathrm{max}}(t-t')} \mathrm{SFR}(t-t') L_{\lambda}^{\mathrm{SSP}}(t', Z(t-t')) \mathrm{d}Z \mathrm{d}t'}{L_{\lambda}(t)}.
\end{equation}

In our radiative transfer calculations, we iteratively solve Equations (\ref{eq:radiativeTransfer}) and (\ref{eq:sourceFunction}) until the fractional change of the source function at all directions on the galaxy surface from one iteration to the next drops below $10^{-10}$.

For comparison with observed photometry at $z=8.312$, we also incorporate intergalactic medium (IGM) attenuation following the model of \citet{inoue2014} to account for the Lyman-alpha break and the stochastic absorption by neutral hydrogen along the line of sight.

\subsubsection{Dust Emission Modeling}

We compute the dust emission spectrum by accounting for both the equilibrium temperature of large grains and the stochastic heating of very small grains and PAHs. The temperature probability distribution, $\mathrm{d}P/\mathrm{d}T$, for grains of each size and species is calculated using Monte Carlo simulations. We adopt the standard methodologies outlined in \citet{fioc2019pegase} for the local energy density estimation and \citet{draine2001infrared, li2001infrared} for the dust heat capacities (including the Debye model and C-H bond modes for PAHs) and cooling rates.

The monochromatic luminosity of a single dust grain with radius $a$ is represented as
\begin{equation}\label{eq:dust_luminosity}
    L_{i}^{\mathrm{grain}} \left( a, \lambda \right) = 4 \pi a^2 \int Q_{\mathrm{abs}}^{i} \left( \lambda \right) B_{\lambda} \left( T \right) \frac{\mathrm{d}P_{i}\left( a \right)}{\mathrm{d}T} \,\mathrm{d}T,
\end{equation}
where $i$ denotes the dust species (silicate, graphite, neutral PAH, or ionized PAH), $B_\lambda(T)$ is the Planck function, $Q_{\mathrm{abs}}^i$ is the absorption efficiency, and $\mathrm{d}P_{i}/\mathrm{d}T$ is the temperature probability distribution derived from the simulation.

Finally, the total luminosity of the galaxy at wavelength $\lambda$ is obtained by integrating over the grain size distribution:
\begin{align}\label{eq:total_luminosity}
    L \left( \lambda \right) & = \sum_{i} \int \frac{\mathrm{d}n_{i}\left( a \right)}{\mathrm{d}a} L_{i}^{\mathrm{grain}} \left( a, \lambda \right) \mathrm{d}a \nonumber \\
    &= 4 \pi \sum_{i} \int \mathrm{d}a \, a^2 \frac{\mathrm{d}n_{i}\left( a \right)}{\mathrm{d}a} \int \mathrm{d}T \, Q_{\mathrm{abs}}^{i}\left( \lambda \right) B_{\lambda} \left( T \right) \frac{\mathrm{d}P_{i}\left( a \right)}{\mathrm{d}T},
\end{align}
where $\mathrm{d}n_{i}\left( a \right)/\mathrm{d}a$ is the size distribution function of dust species $i$ described in Sec. \ref{subsec:DustEvolution}.

\subsubsection{SED of Milky Way like galaxy}
\label{subsubsec:SEDofMW}
\begin{figure}
\centering
\includegraphics[width=0.6\columnwidth]{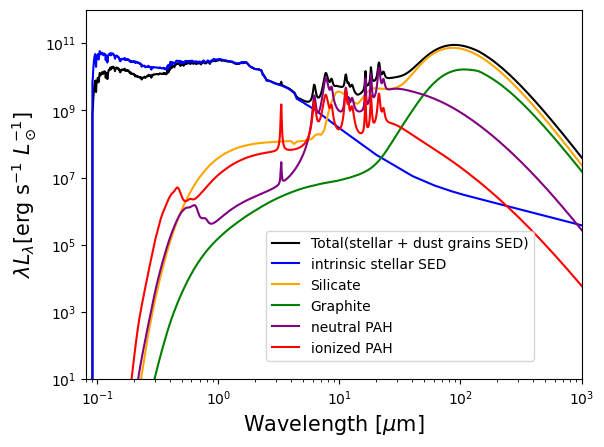}
\caption{SED of a Milky Way-like galaxy at an age of 13\,Gyr computed with the model of \citet{nishida2022new} under the gas infall scenario. The black curve shows the total emission (stellar + dust), the blue curve shows the intrinsic stellar spectrum, and the coloured curves show the contributions from individual dust components (red: ionized PAH; purple: neutral PAH; green: graphite; yellow: silicate).}
\label{image:MW-like-SED}
\end{figure}

To validate the integrated model of dust evolution and radiative transfer, we performed a benchmark test by reproducing the SED of a Milky Way-like galaxy at an age of 13 Gyr using the parameters from \citet{nishida2022new} (Table \ref{table:milkywayParameters}). As shown in Figure \ref{image:MW-like-SED}, the model successfully replicates the characteristic spectral features of a mature galaxy, including the NIR-to-FIR dust emission peaks and PAH features. This confirms that our code accurately tracks the co-evolution of dust size distribution and the resulting radiation field. Having established this baseline, we proceed to apply the model to the extreme environment of MACS0416\_Y1 in the following sections.

\begin{table}
	\centering
	\caption{Milky Way-like galaxy parameters in \citet{nishida2022new}. Here, $M_{\mathrm{gal}}$ is the total baryonic mass, $\tau_{\mathrm{SF}}$ and $\tau_{\mathrm{infall}}$ are the timescales for star formation and gas infall, respectively. $h_{\mathrm{s}}$ represents the scale height of the galaxy disc. The parameters $\eta_{\mathrm{WNM}}$, $\eta_{\mathrm{CNM}}$, and $\eta_{\mathrm{MC}}$ denote the mass fractions of the WNM, CNM, and molecular clouds in the interstellar medium, respectively.}
    \label{table:milkywayParameters}

	\begin{tabular}{|l|ccccccc|} 
		\hline
		$M_{\mathrm{gal}} [\mathrm{M_{\odot}}]$ & $\tau_{\mathrm{SF}} [\mathrm{Gyr}]$ & $\tau_{\mathrm{infall}} [\mathrm{Gyr}]$ & $h_{\mathrm{s}} [\mathrm{pc}]$ & $\eta_{\mathrm{WNM}}$ & $\eta_{\mathrm{CNM}}$ & $\eta_{\mathrm{MC}}$ \\
		\hline
		$10^{11}$ & $3.0$ & $15.0$ & $150$ & $0.5$ & $0.3$ & $0.2$\\
		\hline
	\end{tabular}
\end{table}

\subsection{Target Data: MACS0416\_Y1}
\label{subsec:TargetDataY1}

To investigate the applicability of our model to the early universe, we focus on the galaxy MACS0416\_Y1 at a redshift of $z=8.312$ \citep{tamura2019detections, Bakx2025}. This galaxy is a prime example of the ``dust budget crisis,'' as its observed FIR luminosity suggests a dust mass that traditional chemical evolution models fail to reproduce within the limited cosmic time available ($< 600$ Myr). 

As discussed in Section \ref{sec:introduction}, standard CNM density parameters calibrated for local galaxies ($n_{\mathrm{H,CNM}} \sim 30\,\mathrm{cm}^{-3}$) imply an optically thin geometry, efficient dust destruction by supernova shocks, and a lack of rapid interstellar dust processing. Consequently, such models cannot account for the intense ALMA Band 9 emission \citep{Bakx2025} unless an unrealistic dust yield or an extreme single-temperature assumption is invoked.

Therefore, in the following sections, we explore the need for high-density CNM conditions ($n_{\mathrm{H,CNM}} \sim 7.5 \times 10^3\,\mathrm{cm}^{-3}$). This high-density scenario is consistent with the gravitational collapse and the extreme compactness of star-forming clumps observed in high-redshift environments \citep[e.g.,][]{Shibuya2015, sommovigo2021dust}. By utilizing this framework, we aim to demonstrate that the high optical depth and the suppressed dust destruction in dense environments can resolve the dust budget crisis without invoking extreme dust temperatures.

To ensure a consistent comparison with our model, which focuses on the stellar and dust continuum emission, we corrected the JWST/NIRCam broadband fluxes by subtracting the contributions of strong rest-frame optical emission lines (e.g., [O\,{\sc iii}] and H$\beta$). These contributions were estimated from JWST/NIRSpec spectroscopy, as detailed in Appendix~\ref{sec:appendix_emission_lines}. We exclude the archival Spitzer/IRAC photometry from our analysis. While deblending techniques using high-resolution priors are commonly employed to mitigate source confusion in crowded lensed fields \citep[e.g.,][]{ma2015stellar, zheng2017young}, extracting robust continuum fluxes remains challenging due to the large point spread function (PSF) of IRAC. Furthermore, at $z = 8.31$, the IRAC bands are heavily contaminated by strong rest-optical nebular emission lines. The coupled uncertainties from morphological deblending and emission-line subtraction can introduce artificial flux boosts. Given the availability of superior, high-resolution JWST data, we rely on a combination of high-resolution JWST, HST, and VLT photometry to prevent systematic biases in our physical parameter estimates.

\section{Results}
\label{sec:results}

\subsection{Parameter Exploration and the Dust Budget Crisis}
\begin{figure*}
    \centering
    \begin{minipage}{0.48\textwidth}
        \centering
        \includegraphics[width=\linewidth]{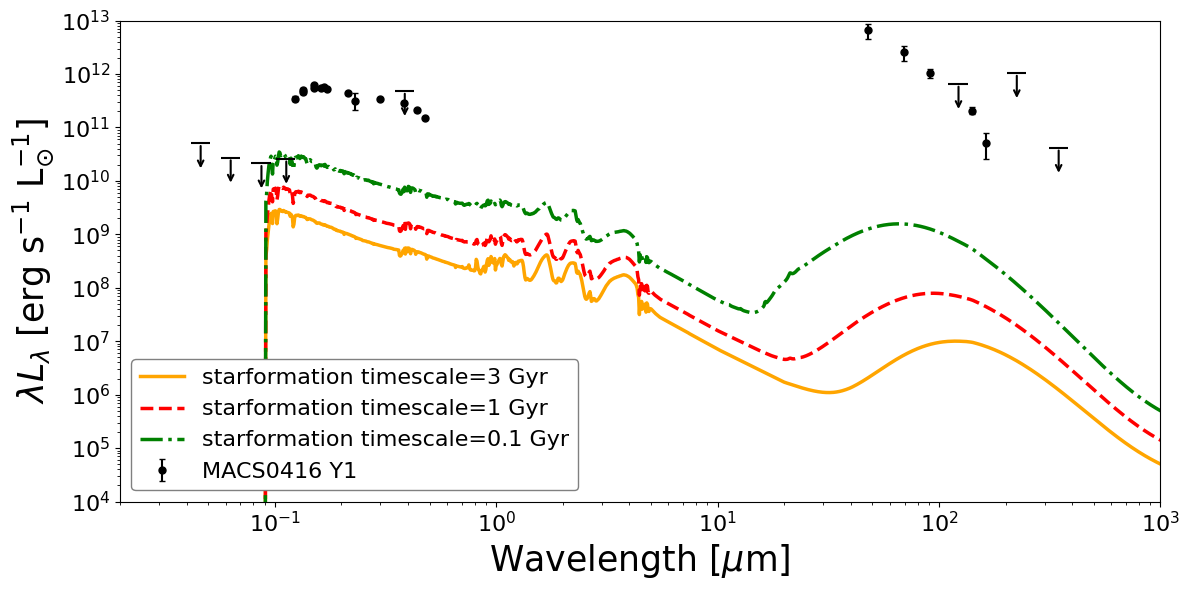}
        \label{fig:timescale_sfr}
    \end{minipage}
    \hfill
    \begin{minipage}{0.48\textwidth}
        \centering
        \includegraphics[width=\linewidth]{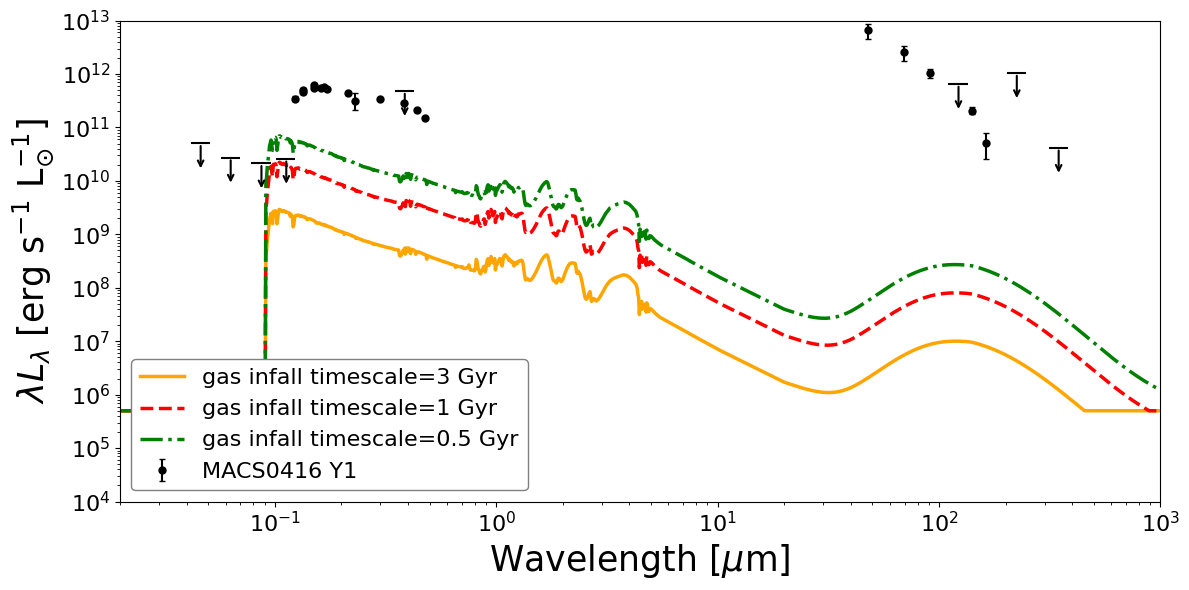}
        \label{fig:timescale_infall}
    \end{minipage}
    \caption{Comparison of the SEDs of a Milky Way-like galaxy with varying evolutionary timescales against the observed data of MACS0416\_Y1 (black markers).
    Left: The curves represent star formation timescales ($\tau_{\rm SF}$) of 3 Gyr (orange), 1 Gyr (red), and 0.1 Gyr (green).
    Right: The curves illustrate gas infall timescales ($\tau_{\rm infall}$) of 3 Gyr (orange), 1 Gyr (red), and 0.5 Gyr (green).
    In both cases, shorter timescales lead to more rapid mass assembly and star formation, significantly enhancing the UV and FIR luminosities. While these adjustments boost the overall intensity, a standard density model fails to fully reproduce the specific shape of the FIR peak.}
    \label{fig:timescale_comparison}
\end{figure*}

We first investigate how the star-formation and gas-infall timescales affect the SED. Figure \ref{fig:timescale_comparison} shows that decreasing the star-formation timescale ($\tau_{\mathrm{SF}}$) from 3.0\,Gyr to 0.1\,Gyr enhances the overall SED, strengthening not only the UV emission due to the intense radiation field but also the FIR dust thermal emission. Furthermore, the shorter $\tau_{\mathrm{SF}}$ intensifies the interstellar radiation field, heating the dust to higher temperatures and shifting the FIR emission peak toward shorter wavelengths. Likewise, the same figure shows that a shorter gas infall timescale ($\tau_{\mathrm{infall}}$) supplies gas more rapidly, primarily boosting the overall normalization of the SED from UV to FIR wavelengths.

Although a model with shorter timescales (e.g., $\tau_{\mathrm{SF}} = 0.1$ Gyr and $\tau_{\mathrm{infall}} = 0.5$ Gyr) can successfully reproduce the UV continuum of MACS0416\_Y1, it results in a total galaxy mass of $\sim 1 \times 10^{11} \mathrm{M}_{\odot}$ at a galaxy age of 500 Myr, which is excessively high for a galaxy at $z \sim 8$. Moreover, despite the enhanced star formation, such standard-density models ($n_{\mathrm{H,CNM}} \sim 30 \mathrm{cm}^{-3}$) fail to match the observed FIR flux, illustrating the persistent ``dust budget crisis.'' This confirms that tuning stellar parameters alone cannot resolve the discrepancy without structural modifications to the ISM.

\subsection{Solution: High-Density CNM and Fixed Clump Radius}
\label{subsec:resolution_high_density}

High-$z$ galaxies are observed to be significantly more compact than local galaxies, suggesting that the ISM density within star-forming regions is substantially higher. We propose that the hydrogen density in the CNM must be as high as $n_{\mathrm{H,CNM}} \sim 7.5 \times 10^3 \mathrm{cm}^{-3}$. This high density is required not primarily to accelerate dust growth, but to ensure sufficient optical depth for UV photons within the compact star-forming regions.

A key refinement in this study is the treatment of the clump radius ($R_{\mathrm{clump}}$). In the previous framework by \citet{nishida2022new}, $R_{\mathrm{clump}}$ was defined by the Jeans length (Equation~\ref{eq:clump_radius}) following \citet{inoue2005attenuation}. However, the Jeans instability criterion fundamentally defines the threshold for the onset of gravitational collapse, not the physical scale of a stable molecular phase. Under this Jeans-length assumption ($R_{\mathrm{clump}} \propto n_{\mathrm{H}}^{-1/2}$), an increase in $n_{\mathrm{H}}$ drastically reduces $R_{\mathrm{clump}}$, preventing the clumps from attaining the optical thickness required to reproduce the observed FIR emission.

Therefore, decoupling the clump size from the Jeans condition is physically motivated for modeling
the extreme environments of high-$z$ starbursts. We treat $R_{\mathrm{clump}}$ as a fixed physical
parameter and adopt $R_{\mathrm{clump}} = 10.4$ pc, consistent with the typical scale of
star-forming complexes \citep[e.g.,][]{Larson1981}. By fixing the radius and increasing the density to
$n_{\mathrm{H}} = 7.5 \times 10^3\,\mathrm{cm}^{-3}$, the clumps become sufficiently optically thick
to stellar UV/optical photons. This compact geometry efficiently traps UV photons and boosts
the FIR output via IR re-radiation, providing a primary solution to the ``dust budget crisis''.
We quantify the wavelength dependence of $\tau_{\rm cl}(\lambda)$ and $P_{\rm esc}(\lambda)$ for the
best-fit model in Appendix~\ref{app:tau_pesc} (Fig.~\ref{fig:tau_pesc_n0}). Although rapid dust mass growth via accretion in such high-density environments also contributes, the enhanced optical depth in the UV and optical bands driven by this compact geometry is the dominant factor in reproducing the extreme FIR luminosity.

\subsection{Statistical Parameter Estimation via MCMC}
\label{subsec:MCMCRESULT}
While we formulated the baseline galaxy evolution using an exponentially declining gas infall rate to validate the local benchmark, high-redshift galaxies at $z > 6$ are expected to undergo a rapid mass assembly phase. To naturally reproduce this, we replaced the standard infall model with a delayed exponential infall model for the fitting of MACS0416\_Y1 \citep{Takeuchi2026}:
\begin{equation}
\label{eq:risingSFH}
    \frac{dM_{\rm infall}}{dt} = M_{\rm infall} \frac{t}{\tau_{\rm infall}^2} \exp\left(-\frac{t}{\tau_{\rm infall}}\right).
\end{equation}
The accretion rate peaks at $t \simeq \tau_{\rm infall}$. 
When combined with the Schmidt law (Equation~\ref{eq:schmidtLaw}), the resulting star formation history initially rises. 
Because star formation is regulated by the accumulated ISM gas mass, the peak of the SFR does not necessarily coincide with the peak of the gas infall rate. 
Instead, the peak time of the SFR depends on parameters such as the star formation timescale $\tau_{\rm SF}$ and the treatment of outflows or mass return.

To determine the best-fitting physical parameters, we performed a Markov Chain Monte Carlo (MCMC) analysis using the affine-invariant ensemble sampler \textsc{emcee} \citep{ForemanMackey2013}. 
The likelihood function was defined in logarithmic space to ensure numerical stability. We accounted for asymmetric error bars in the observed fluxes for detections. For non-detections (upper limits), we applied a penalty term to the likelihood only when the model flux exceeded the observed upper limit.

To efficiently explore the parameter space, we first pre-computed a comprehensive library of $100{,}000$ SED templates. The parameter grid spans 20 logarithmically spaced bins for the CNM density parameter ($n_{0, \mathrm{CNM}} \in [10^3, 10^6]\,\mathrm{cm}^{-3}$) and 5 discrete galaxy ages ($t_{\mathrm{gal}} \in [40, 80]\,\mathrm{Myr}$ in steps of $10\,\mathrm{Myr}$). For the star formation history and mass parameters, we generated $10 \times 10 \times 10$ logarithmically spaced grid points: total infalling baryon mass ($M_{\mathrm{total}} \in [5\times10^{10}, 5\times10^{11}]\,M_{\odot}$), star formation timescale ($\tau_{\mathrm{SF}} \in [10^7, 5\times10^8]\,\mathrm{yr}$), and gas infall timescale ($\tau_{\mathrm{infall}} \in [10^7, 5\times10^8]\,\mathrm{yr}$). 

Leveraging this extensive pre-computed grid, we performed a grid-interpolated MCMC analysis. For each combination of the discrete parameters ($n_{0, \mathrm{CNM}}$ and $t_{\mathrm{gal}}$), we ran MCMC chains to sample the continuous parameters ($M_{\mathrm{total}}$, $\tau_{\mathrm{SF}}$, $\tau_{\mathrm{infall}}$), assuming uniform priors in linear space within their respective boundaries. The sampling was executed using 32 walkers for 1000 steps, discarding the initial 10\% of the chains as burn-in to ensure convergence. The global best-fit parameter set was ultimately determined by the maximum log-likelihood across all grid combinations.

\begin{figure}
\centering
\includegraphics[width=0.9\columnwidth]{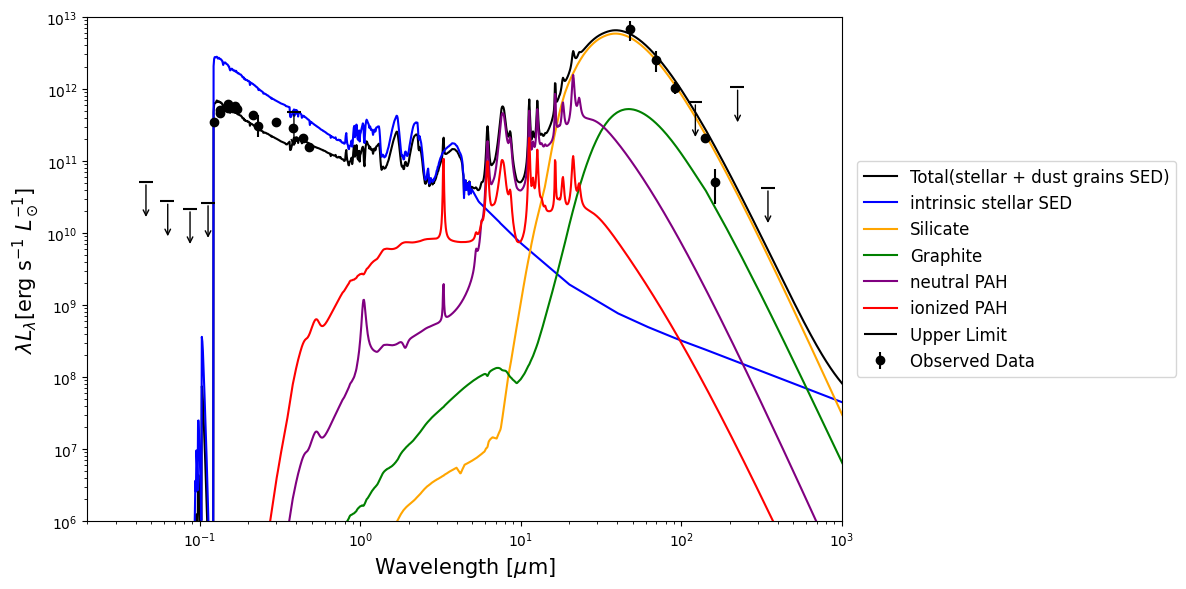}
\caption{Best-fit SED of MACS0416\_Y1 at $z=8.312$. Black markers with error bars represent the observational data \citep{tamura2019detections, Ma_2024, Bakx2025}. The total model SED (black line) reproduces the observed UV and FIR fluxes. The silicate component (orange line) provides the dominant contribution around the FIR peak. In our compact, high-density ISM model, the high optical depth and the size-resolved (multi-temperature) dust population enable a consistent fit without invoking a globally extreme single dust temperature (see Figure~\ref{image:TemperatureDistribution}).}
\label{image:BestFitSED}
\end{figure}

The posterior distributions yield the best-fit parameters summarized in Table \ref{table:distant_galaxy_parameters}. Figure \ref{image:BestFitSED} shows the resulting SED, which provides an excellent fit to the observational data. Notably, the analysis confirms that a high-density scenario ($n_{\mathrm{H,CNM}} \approx 7.5 \times 10^3\,\mathrm{cm}^{-3}$) is statistically preferred. At the best-fit galaxy age of 80\,Myr, the inferred stellar mass $M_{*} = 1.13 \times 10^{10}\,\mathrm{M}_{\odot}$ and dust mass $M_{\mathrm{d}} = 6.02 \times 10^{7}\,\mathrm{M}_{\odot}$ self-consistently characterize the rapid dust assembly and the compact starburst nature of this $z \sim 8$ system. Although our best-fit model prefers an age of $80\mathrm{\,Myr}$, younger templates combined with sufficiently high CNM densities yield statistically indistinguishable fits (see Figure \ref{fig:delta_lnL} in Appendix \ref{sec:appendix_grid_search} for details on the parameter degeneracy). This indicates that a high CNM density is the fundamental requirement to reproduce the observations.

\begin{table}
	\centering
	\caption{Best-fit parameters for MACS0416\_Y1. The upper section lists the parameters and their $1\sigma$ uncertainties derived from the MCMC posterior distributions. The lower section lists the galaxy age ($t_{\mathrm{gal}}$) and CNM density parameter ($n_{0, \mathrm{CNM}}$), which were determined via a discrete grid search to minimize the log-likelihood.}
	\label{table:distant_galaxy_parameters}
	\begin{tabular}{lcr} 
		\hline
		Parameter & Physical Quantity & Value\\
		\hline
        \multicolumn{3}{c}{MCMC Posterior Estimates} \\
		\hline
		$M_{\mathrm{total}}$ & Total infalling baryon mass & $(2.67 \pm 1.26) \times 10^{11}\,\mathrm{M}_{\odot}$ \\
		$\tau_{\mathrm{SF}}$ & Star formation timescale & $(1.54 \pm 1.32) \times 10^8\,\mathrm{yr}$\\
        $\tau_{\mathrm{infall}}$ & Gas infall timescale & $(7.91 \pm 5.45) \times 10^7\,\mathrm{yr}$\\
        \hline
        \multicolumn{3}{c}{Grid Search Point Estimates} \\
        \hline
        $t_{\mathrm{gal}}$ & Galaxy age & $80\,\mathrm{Myr}$\\
		$n_{0, \mathrm{CNM}}$ & CNM density parameter & $2 \times 10^5\,\mathrm{cm}^{-3}$\\
		\hline
	\end{tabular}
\end{table}

\subsection{Physical Properties of the Best-fit Model}
\label{subsec:physical_properties}

To understand the physical mechanisms behind our successful reproduction of the MACS0416\_Y1 SED, we analyze the internal properties of our best-fit model, focusing on the dust size distribution, chemical composition, and temperature distribution.

\begin{figure}
\centering
\includegraphics[width=1.0\columnwidth]{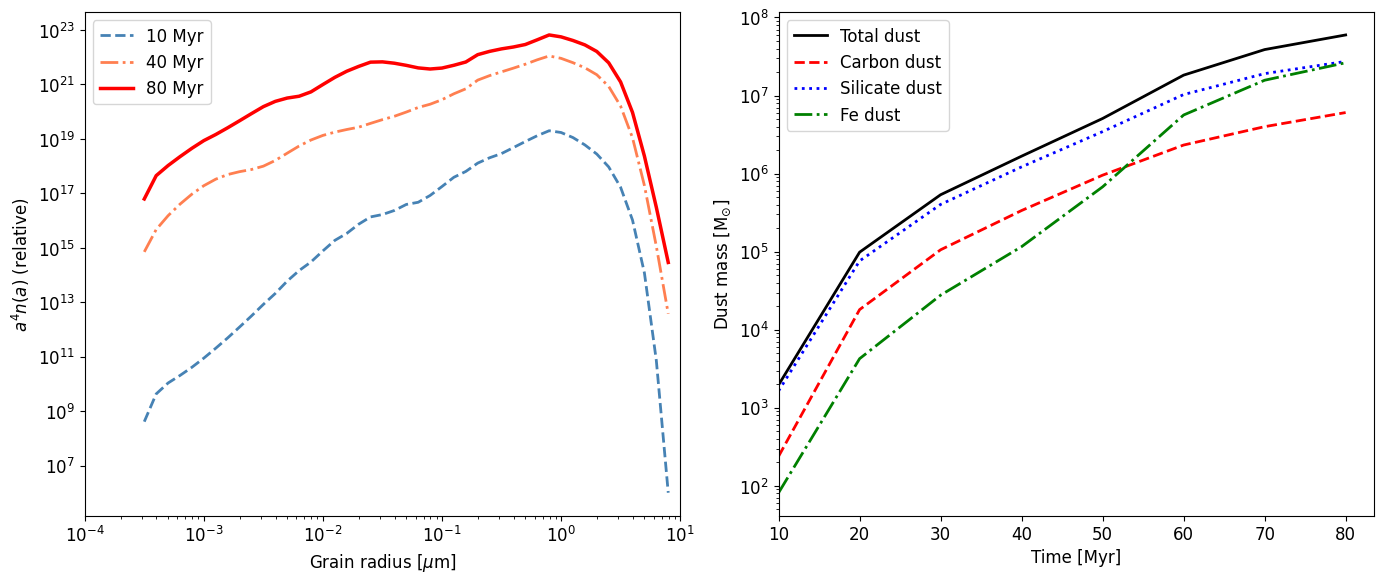}
\caption{Left: Time evolution of the dust size distribution at 10\,Myr (blue dashed), 40\,Myr (orange dot-dashed), and 80\,Myr (red solid). The prominent increase in small-to-medium grains clearly reflects the shattering-triggered accretion phase. Right: Evolution of total dust mass (black solid) and its composition: Carbon (red dashed), Silicate (blue dotted), and Iron (green dot-dashed). After an initial phase governed by stellar injection ($\lesssim 40$\,Myr), interstellar accretion rapidly dominates the mass assembly, primarily driven by Silicate and Iron grains.}
\label{image:TotalDustMassEvolution}
\end{figure}

\begin{figure}
\centering
\includegraphics[width=1.0\columnwidth]{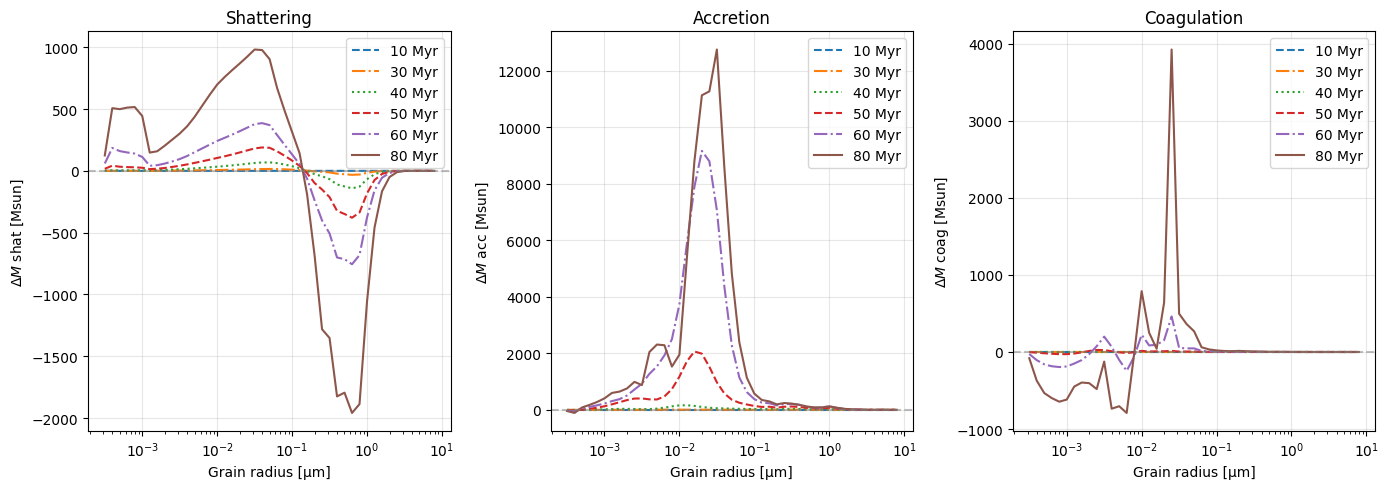}
\caption{Dust mass change rate ($\Delta M$) driven by shattering (left), accretion (middle), and coagulation (right) as a function of grain radius. The curves represent different galactic ages from 10 to 80\,Myr. A clear sequential evolution is observed: shattering of large grains ($a > 0.1\,\mu\mathrm{m}$) becomes prominent around 40-50\,Myr (green dotted line), supplying a massive number of small grains. This critical supply of total surface area acts as a trigger for an explosive accretion phase, dominating the mass growth around $a \sim 10^{-2}\,\mu\mathrm{m}$ after 50\,Myr (red dashed line). At later stages ($\gtrsim 60$\,Myr, purple dot-dashed and brown solid lines), coagulation begins to efficiently merge the smallest grains ($a < 10^{-2}\,\mu\mathrm{m}$) into medium-sized grains. This shattering-triggered accretion sequence is the fundamental driver of the rapid dust mass assembly in the high-density environment.}
\label{image:DustEvolutionRates}
\end{figure}

\subsubsection{Interstellar Processing and Dust Size Evolution}
\label{subsubsec:dustSizeAndEvolution}
The left panel of Figure \ref{image:TotalDustMassEvolution} shows the time evolution of the dust size distribution in the high-density environment. Because the ordinate is $a^{4}n(a)$, it traces the dust mass distribution per logarithmic size interval. At $10\,\mathrm{Myr}$, the mass distribution is weighted toward large grains ($a \sim 0.5$--$2\,\mu\mathrm{m}$), consistent with the direct injection spectrum from stellar sources. As the system evolves, the mass distribution at small-to-intermediate sizes ($a \sim 0.01$--$0.1\,\mu\mathrm{m}$) increases substantially.

The physical origin of this evolution is clarified by the process-dependent mass-transfer rates in Figure \ref{image:DustEvolutionRates}. By $\sim 50\,\mathrm{Myr}$, shattering becomes clearly active and transfers mass from large grains ($a \gtrsim 0.1\,\mu\mathrm{m}$) to smaller sizes, increasing the total grain surface area. This enhanced surface area facilitates rapid grain growth by accretion, which provides the dominant net dust-mass increase at later times, with the strongest growth occurring at $a \sim \mathrm{few}\times 10^{-2}\,\mu\mathrm{m}$. In the late stage ($\gtrsim 60\,\mathrm{Myr}$), coagulation becomes significant and transfers mass from the smallest grains to intermediate sizes ($a \sim 10^{-2}$--$10^{-1}\,\mu\mathrm{m}$). Overall, multiple processes operate concurrently in the high-density ISM; shattering and coagulation primarily redistribute dust mass across sizes, whereas accretion dominates the net mass growth while reshaping the size distribution.

\subsubsection{Grain Composition and Rapid Mass Accumulation}
\label{subsubsec:grainComposition}
The right panel of Figure \ref{image:TotalDustMassEvolution} presents the evolution of the total dust mass and its compositional breakdown. During the early phase ($\lesssim 40$--$50\,\mathrm{Myr}$), accretion is still weak in Figure \ref{image:DustEvolutionRates}; accordingly, the increase in total dust mass is dominated by the stellar injection term (primarily supernovae) in our model. Once accretion becomes efficient at later times, interstellar dust growth dominates the mass accumulation. In this phase, silicate and Fe grains provide the main contribution to the rapid rise of the total dust mass, while carbon dust remains subdominant.

This interstellar dust assembly complements our structural argument. While the compact ISM geometry ensures the high optical depth required to trap UV photons and boost the FIR output (Section \ref{subsec:resolution_high_density}), the large dust mass itself is rapidly built up through the co-evolution of the dust size distribution and interstellar processing. Taken together, the shattering-enabled rapid accretion and the compact environment provide a physically motivated pathway that can alleviate the ``dust budget crisis'' in high-$z$ starbursts.

\subsubsection{Temperature Distribution and FIR Emission}
\label{subsubsec:tempdist}
\begin{figure}
\centering
\includegraphics[width=0.5\columnwidth]{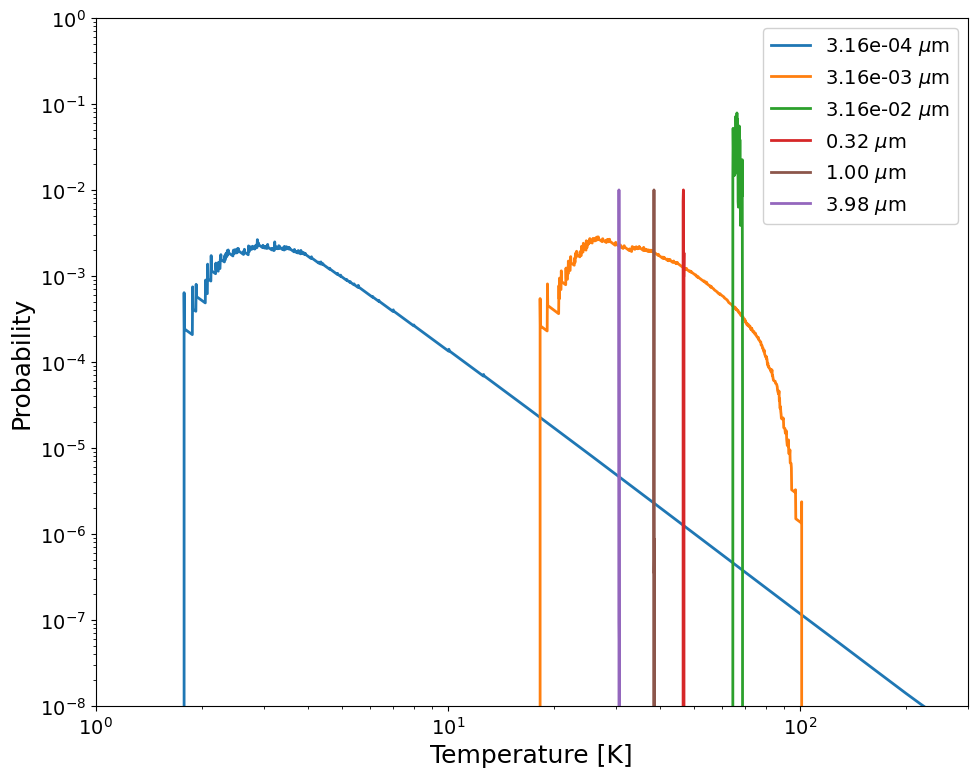}
\caption{Binned temperature probability distribution of silicate grains for radii
$3.16 \times 10^{-4}$ to $3.98\,\mu\mathrm{m}$ computed with our Monte Carlo
stochastic-heating calculation. The distribution is normalized such that $\sum_i P_i = 1$,
where $P_i$ is the probability in the $i$-th temperature bin.
Very small grains show broad non-equilibrium distributions with low-probability high-$T$ tails,
while grains with $a \gtrsim 10^{-2}\,\mu\mathrm{m}$ exhibit narrow distributions close to equilibrium,
with characteristic temperatures of $\sim 30$--$70\,\mathrm{K}$ in our model.
For the largest grains, the distribution is narrower than the temperature binning and is therefore
shown as a vertical line at the equilibrium temperature $T_{\rm eq}$.}
\label{image:TemperatureDistribution}
\end{figure}

Figure \ref{image:TemperatureDistribution} shows the binned temperature probability distribution
of silicate grains computed with our Monte Carlo stochastic-heating calculation. The distribution is
normalized such that the sum of probabilities over all temperature bins equals unity,
$\sum_i P_i = 1$, where $P_i$ is the probability that a grain resides in the $i$-th temperature bin.
Small grains have low heat capacities and therefore exhibit broad, non-equilibrium temperature
distributions, whereas sufficiently large grains approach a well-defined equilibrium temperature with
a very narrow distribution.

For grains large enough to be close to thermal equilibrium ($a \gtrsim 10^{-2}\,\mu\mathrm{m}$),
the distributions are sharply peaked (near-equilibrium), with characteristic temperatures of
$\sim 30$--$70\,\mathrm{K}$ depending on grain size in our radiation field. In contrast, very small
grains ($a \lesssim 10^{-3}\,\mu\mathrm{m}$) are stochastically heated: they spend most of the time at
low temperatures (a few--$\sim 10\,\mathrm{K}$) with a low-probability high-temperature tail extending
to $\gtrsim 10^{2}\,\mathrm{K}$.

Compared to the single-temperature assumption of $\sim 91\,\mathrm{K}$ adopted in previous work
\citep{Bakx2025}, our result indicates that reproducing the ALMA Band~9 flux does not require
extreme dust temperatures derived from a single-temperature treatment. Instead, the intense FIR emission is primarily driven by
efficient trapping of stellar UV/optical photons in compact, high-$\tau$ clumps and subsequent IR
re-radiation, together with the rapid build-up of the dust mass discussed in
Figure~\ref{image:TotalDustMassEvolution}. In this framework, the dominant Band~9 emission arises
primarily from near-equilibrium grains in the intermediate-size range
($a\simeq 0.01$--$0.1\,\mu\mathrm{m}$), while very small grains contribute mainly via stochastic
heating and a low-probability high-$T$ tail.

\subsubsection{Grain-size origin of the warm IR peak and ALMA emission}
\label{subsubsec:grainsize_origin}

\begin{figure*}
    \centering
    \includegraphics[width=\textwidth]{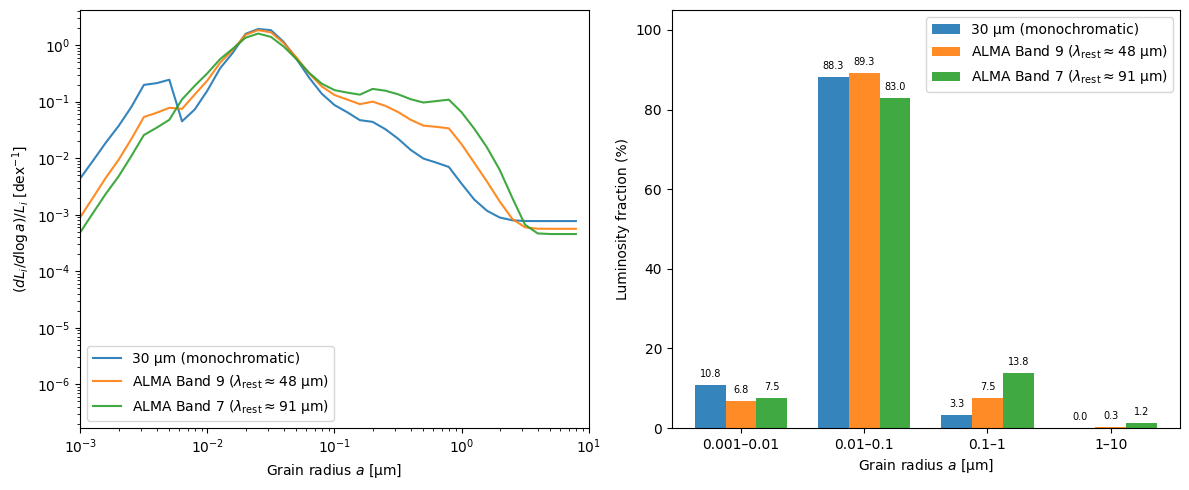}
\caption{Grain-size origin of the warm IR peak and the ALMA continua in the best-fit model at
$t_{\rm gal}=80\,\mathrm{Myr}$ (emergent emission; escape included).
\textit{Left:} Normalized size-resolved contributions, $(dL_i/d\log a)/L_i$, for the monochromatic
emission at $30\,\mu$m and for the band-integrated ALMA Band~9 and Band~7 continua (rest-frame
$\lambda\simeq 48$ and $91\,\mu$m at $z=8.312$), where each curve is normalized such that
$\int d\log a\, (dL_i/d\log a)/L_i = 1$.
\textit{Right:} Cumulative luminosity fractions integrated over non-overlapping grain-size bins.
Intermediate-size grains ($a=0.01$--$0.1\,\mu$m) dominate the $30\,\mu$m emission and the Band~9
continuum (88.3\% and 89.3\%, respectively), while the relative contribution from larger grains
increases toward longer wavelengths (e.g.\ Band~7).}
    \label{fig:grainsize_origin}
\end{figure*}

To clarify which grain sizes dominate the emission at the warm IR peak and in the ALMA continuum,
we compute the size-resolved luminosity contribution, $dL_i/d\log a$, for the monochromatic emission
at $\lambda=30\,\mu$m and for the band-integrated ALMA Band~9 and Band~7 continua.
Figure~\ref{fig:grainsize_origin} (left) shows the normalized distributions $(dL_i/d\log a)/L_i$
($\int d\log a\, (dL_i/d\log a)/L_i=1$), highlighting the grain sizes that dominate the emission,
while Figure~\ref{fig:grainsize_origin} (right) summarizes the corresponding cumulative luminosity
fractions in non-overlapping grain-size bins.

We find that intermediate-size grains with $a=0.01$--$0.1\,\mu$m dominate both the $30\,\mu$m emission
and the Band~9 continuum, contributing 88.3\% and 89.3\% of the emergent luminosity, respectively.
The normalized contribution peaks at $a\simeq 0.03\,\mu$m, consistent with the near-equilibrium
temperature distributions in Figure~\ref{image:TemperatureDistribution}. Toward longer wavelengths,
the relative contribution from larger grains increases: in Band~7 the $a=0.1$--$1\,\mu$m bin contributes
13.8\%, compared to 7.5\% in Band~9.

\section{Discussion}
\label{sec:discussion}
\subsection{Resolving the Dust Budget Crisis via Geometric Effects and Enhanced Survival}
\label{subsec:discussion_dust_budget}
A key result of this study is that the observed SED of MACS0416\_Y1 can be reproduced in a high-density CNM environment ($n_{\mathrm{H,CNM}} \simeq 7.5 \times 10^{3}\,\mathrm{cm^{-3}}$), for which our best-fit model yields a dust mass $M_{\mathrm{d}} = 6.02 \times 10^{7}\,\mathrm{M}_{\odot}$ at $t_{\mathrm{gal}} = 80\,\mathrm{Myr}$ (Table~\ref{table:distant_galaxy_parameters}; Figure~\ref{image:BestFitSED}).
The build-up of this dust mass proceeds in two phases. During the early stage ($\lesssim 40$--$50\,\mathrm{Myr}$), the increase is dominated by stellar injection (primarily SNe), while interstellar accretion is still inefficient. At later times, once shattering has increased the total grain surface area, accretion in the dense ISM becomes efficient and dominates the net dust-mass growth (Figures~\ref{image:TotalDustMassEvolution}--\ref{image:DustEvolutionRates}). Thus, the large dust mass is not achieved by invoking ad hoc yields, but emerges self-consistently from the co-evolution of the grain-size distribution and interstellar processing. This signifies a rapid transition from stellar-injection-dominated to ISM-processing-dominated growth, implying that dust properties in such compact $z \sim 8$ systems are not a direct imprint of stellar yields alone, but are set by the coupled evolution of grain sizes and ISM conditions.

In addition, the high-density scenario alleviates the ``dust budget crisis'' through two coupled effects. First, the compact, optically thick geometry increases the conversion efficiency of stellar UV radiation into FIR emission by trapping photons and re-radiating them in the infrared. Second, the efficiency of SN shock destruction is reduced because the swept-up mass decreases with increasing ambient density (e.g., $M_{\mathrm{swept}} \propto n_{\mathrm{SN}}^{-0.202}$), improving the survival probability of newly formed grains. We note, however, that the shock-destruction prescription is extrapolated beyond its original calibration range (see Section~\ref{subsec:limitations}). Overall, our results suggest that the apparent ``crisis'' in compact high-$z$ systems stems from a combination of underestimated optical depth and misestimated dust survival/growth efficiencies in dense early-universe environments.

Our best-fit model yields a relatively older galaxy age ($\approx$ 80 Myr) and a larger stellar mass ($M_{*} = 1.13 \times 10^{10}\,\mathrm{M}_{\odot}$) for MACS0416\_Y1. In contrast, previous studies relying on rest-UV/optical SED fitting have reported significantly younger ages and lower masses (e.g., Age $\approx$ 5 Myr, $M_* \approx 10^9 M_{\odot}$; \citet{Ma_2024}). This discrepancy is primarily driven by the outshining effect, where the recent starburst dominates the UV/optical emission, masking the underlying older stellar populations. However, as qualitatively pointed out by \citet{Bakx2025}, a purely young stellar population ($\sim$ 4–5 Myr) physically cannot produce the observed massive dust reservoir via supernovae and AGB stars; the contribution from an older stellar component is strictly required. Because our framework self-consistently solves the UV-to-FIR energy balance coupled with dust chemical evolution, it quantitatively captures this hidden older stellar component and the cumulative star formation history. Consequently, our model naturally yields an older age and a correspondingly higher stellar mass, providing robust quantitative support for the interpretation by \citet{Bakx2025}.

\subsection{Physical Consistency of Dust Temperature and ISM Geometry}
\label{subsec:discussion_temp_geometry}
Our model reproduces the ALMA Band~9 flux with moderate equilibrium temperatures ($30$--$70\,\mathrm{K}$) by accounting for size-resolved dust emission. While the bulk dust mass resides in large grains ($a \sim 1\,\mu\mathrm{m}$) at $T_{\rm eq} \sim 35$--$45\,\mathrm{K}$, the warm IR peak ($\lambda_{\rm rest} \sim 30\,\mu\mathrm{m}$) and the Band~9 emission are instead dominated by warmer, intermediate-size grains ($a = 0.01$--$0.1\,\mu\mathrm{m}$; $T_{\rm eq} \sim 70\,\mathrm{K}$). These grains contribute $\sim 89\%$ of the emission in both regimes, whereas the mass-dominant population contributes less than $8\%$ (Figures~\ref{image:TemperatureDistribution}, \ref{fig:grainsize_origin}). This size-dependent emission profile is a direct consequence of our evolution model: shattering and subsequent accretion in the dense ISM boost the mass reservoir of intermediate grains (Figures~\ref{image:TotalDustMassEvolution}, \ref{image:DustEvolutionRates}). Thus, the observed Band~9 flux is naturally explained by this self-consistently grown grain population, eliminating the need for extreme dust temperatures derived from a single-temperature treatment..

The key difference from previous frameworks (\citealt{inoue2005attenuation}; \citealt{nishida2022new}) arises from the treatment of the clump radius ($R_{\mathrm{clump}}$). By decoupling $R_{\mathrm{clump}}$ from the density-dependent Jeans-length scaling and adopting a fixed physical scale ($R_{\mathrm{clump}}=10.4\,\mathrm{pc}$), we increase the dust surface density and enhance UV photon trapping. Furthermore, driven by the peak gas infall around $t \sim 80\,\mathrm{Myr}$, our star formation history (SFH) is in a rapidly rising phase. This actively star-forming period provides a strong UV radiation field that enhances dust heating. Shortening the SFH timescale further strengthens this field, shifting the dust emission peak toward shorter wavelengths. In summary, the physical consistency between our grain evolution model and the dense ISM geometry provides a robust explanation for the warm IR peak without invoking extreme global conditions.

\subsection{Other Potential Dust Sources and Material Effects}While our model focuses on AGB stars, SNe II, and grain growth, other contributors such as SNe Ia or Wolf-Rayet stars may also play minor roles. However, SNe Ia are generally expected to contribute negligibly to the total budget at these early epochs \citep{nozawa2011formation}. Additionally, the adoption of amorphous dust properties instead of crystalline structures could affect the extinction curves \citep{asano2014evolution}. Potential AGN activity could also boost the mid-IR continuum, though for MACS0416\_Y1, the emission appears spatially extended, suggesting that an AGN is unlikely to dominate the FIR unless it is extremely luminous \citep{Bakx2025, Mullaney2011}.\subsection{Limitations of the Current Model}\label{subsec:limitations}We note that the formulae for $M_{\rm swept}$ were originally calibrated for lower density environments ($n_{\rm SN} \le 30~{\rm cm}^{-3}$). While extrapolating this to our extreme high-density regime captures the physical expectation, quantitative uncertainties remain regarding the exact efficiency of shock destruction. Future hydrodynamical simulations of SN shocks in extremely dense molecular clouds are required to confirm these environmental effects.

We note that the cosmic microwave background (CMB) temperature at $z=8.312$ is $T_{\rm CMB} \approx 25.4 \, \rm K$. We verified that setting this as a strict lower limit for the dust temperature distribution has a negligible impact on our resulting SED and derived physical properties, as the emission is dominated by grains at much higher characteristic temperatures ($\sim 30-70 \, \rm K$).

\section{Conclusions}

In this study, we extended the dust evolution model and the SED framework to the high-redshift universe, focusing on the galaxy MACS0416\_Y1 at $z=8.312$.
Our main conclusions are summarized as follows:

\begin{enumerate}
    \item We successfully reproduce the observed UV-to-FIR SED of MACS0416\_Y1 with a compact, high-density CNM environment.
    The best-fit solution prefers an extreme CNM density of $n_{\mathrm{H,CNM}}\approx 7.5\times 10^{3}\,\mathrm{cm^{-3}}$ (corresponding to $n_{0,\mathrm{CNM}}=2\times 10^{5}\,\mathrm{cm^{-3}}$) at a galaxy age of $t_{\mathrm{gal}}=80\,\mathrm{Myr}$, yielding a dust mass of $M_{\mathrm{d}}=1.08\times 10^{8}\,M_{\odot}$.

    \item The high FIR output does not require a globally extreme single dust temperature inferred from optically thin, single-temperature fits.
Our size-resolved treatment predicts a multi-temperature dust population, in which grains have equilibrium (or near-equilibrium) temperatures of
$\sim 30$--$70\,\mathrm{K}$ depending on size: the mass-dominant largest grains remain at $\sim 35$--$45\,\mathrm{K}$, while intermediate-size grains
reach higher characteristic temperatures ($\sim 70\,\mathrm{K}$) and the very smallest grains exhibit a low-probability high-$T$ tail extending to
$\gtrsim 10^{2}\,\mathrm{K}$. Importantly, the emission near the IR peak ($30\,\mu\mathrm{m}$) and the band-integrated ALMA Band~9 continuum
(rest-frame $48\,\mu\mathrm{m}$) is dominated by intermediate-size grains with $a=0.01$--$0.1\,\mu\mathrm{m}$, contributing 88.3\% and 89.3\% of the
emergent luminosity, respectively.

    \item The solution to the ``dust budget crisis'' arises from coupled effects of geometry and dust processing in a dense ISM.
    Fixing the clump radius at $R_{\mathrm{clump}}=10.4\,\mathrm{pc}$ (decoupled from the Jeans-length scaling) increases the dust surface density and hence the optical depth in the UV and optical bands,
    efficiently trapping stellar UV photons and boosting the FIR emission.
    In addition, dust mass growth proceeds in two phases: stellar injection dominates at early times, whereas shattering increases the total surface area and triggers efficient accretion,
    which dominates the net dust-mass growth at later times, with coagulation becoming non-negligible in the late stage.

    \item The high-density environment may also reduce the efficiency of SN shock destruction through the reduced swept-up mass
    (e.g., $M_{\rm swept} \propto n_{\rm SN}^{-0.202}$ in our adopted prescription). However, this prescription is extrapolated beyond its original calibration range,
    and the quantitative destruction efficiency in such extreme densities remains uncertain.
\end{enumerate}

Overall, our results suggest that the dust budget crisis in compact $z\sim 8$ systems can be alleviated by the combined effects of high optical depth
in dense clumps (boosting UV-to-FIR conversion) and a ``compressed'' dust evolution in which shattering-enabled accretion rapidly builds up the dust mass
and yields a multi-temperature grain population, with the warm IR peak and Band~9 continuum being dominated by intermediate-size grains
($a=0.01$--$0.1\,\mu\mathrm{m}$).

\section*{Acknowledgments}
This work has been supported by the Japan Society for the Promotion of Science (JSPS) Grants-in-Aid for Scientific Research (21H01128 and 24H00247). 
This work has also been supported in part by the Sumitomo Foundation Fiscal 2018 Grant for Basic Science Research Projects (180923), and the Collaboration Funding of the Institute of Statistical Mathematics ``Machine-Learning-Based Cosmogony: From Structure Formation to Galaxy Evolution''. This work was financially supported by JST SPRING, Grant Number
JPMJSP2125.The author RRK would like to take this opportunity to thank the
“THERS Make New Standards Program for the Next Generation Researchers. MH is supported by JSPS KAKENHI Grant No. 22H04939.

\section*{Data Availability}
The data and code underlying this article will be shared on reasonable request to the corresponding author.


\bibliographystyle{mnras}
\bibliography{example} 




\appendix
\section{Emission Line Corrections for JWST/NIRCam Photometry}
\label{sec:appendix_emission_lines}

To accurately constrain the continuum emission and dust properties of MACS0416\_Y1, we corrected the JWST/NIRCam broadband photometry for rest-frame optical emission line contamination ($3000$--$5000$\,\AA). The emission line fluxes were extracted from JWST/NIRSpec IFS observations (Program ID: 1208, PI: C. Willott; total exposure time $18{,}206$\,s). The data were reduced using the JWST Pipeline version 1.13.4 with CRDS context 1193.

We measured the total system line fluxes using a curve-of-growth analysis centered on the centroid of the integrated intensity map. For most lines, integrating within the $2\sigma$ region of the [O\,{\sc iii}] $\lambda5008$ moment-0 map successfully recovered the total flux. For the [O\,{\sc iii}] $\lambda5008$ and [O\,{\sc ii}] doublets, a 10\% aperture correction was applied based on the curve-of-growth.

We compared these line fluxes to the total NIRCam photometry reported by \citet{Ma_2024}. As a first-order approximation, we assumed a top-hat filter response, estimating the continuum flux within the full width at half maximum (FWHM) of each filter. The estimated fractional contributions of the emission lines to the observed broadband fluxes are as follows:
\begin{itemize}
    \item \textbf{F356W}: $\sim 26$\% (contributing lines: [O\,{\sc ii}], H9, [Ne\,{\sc iii}] $\lambda3869$, H8, H$\epsilon$, H$\delta$)
    \item \textbf{F410M}: $\sim 16$\% (contributing lines: H$\gamma$)
    \item \textbf{F444W}: $\sim 75$\% (contributing lines: H$\gamma$, H$\beta$, [O\,{\sc iii}] $\lambda\lambda4960, 5008$)
\end{itemize}

\section{Degeneracy in the Parameter Space}
\label{sec:appendix_grid_search}

This appendix provides a supplementary visualization of the grid search results to illustrate the parameter degeneracy between the template age and the CNM density.

\begin{figure}
    \centering
    \includegraphics[width=0.6\columnwidth]{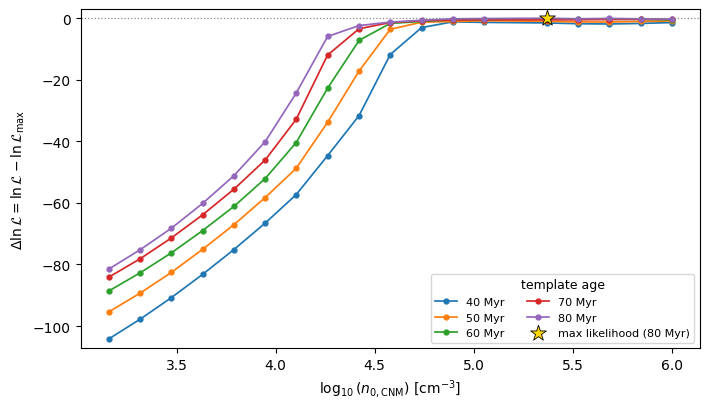}
    \caption{The relative log-likelihood ($\Delta \ln \mathcal{L}$) as a function of the CNM density ($n_{0,\mathrm{CNM}}$) for different template ages. While the global maximum is found at $80\mathrm{\,Myr}$, younger templates combined with sufficiently high CNM densities yield statistically indistinguishable fits ($\Delta \ln \mathcal{L} \sim 0$). This demonstrates that a sufficiently high CNM density is the fundamental requirement to reproduce the observed SED, rather than a unique template age.}
    \label{fig:delta_lnL}
\end{figure}

\section{Wavelength-dependent UV photon trapping in dust clumps}
\label{app:tau_pesc}

In this Section, we quantify how the clump optical depth and the corresponding photon escape
probability depend on wavelength, and how strongly they change with the normalization parameter
$n_{0,\mathrm{CNM}}$ of the effective two-phase pressure relation (Eq.~\ref{eq:pressure_equilibrium}).
For a given model snapshot, we evaluate the wavelength-dependent clump optical depth
$\tau_{\rm cl}(\lambda)$ (Eq.~\ref{eq:optical_depth_clump}) and the associated escape probability
$P_{\rm esc}(\lambda)$ (Eq.~\ref{eq:photonEscapeProbability}), which together determine how efficiently
stellar UV/optical photons are trapped inside compact star-forming clumps before being re-radiated in
the infrared.

\begin{figure*}
    \centering
    \includegraphics[width=\textwidth]{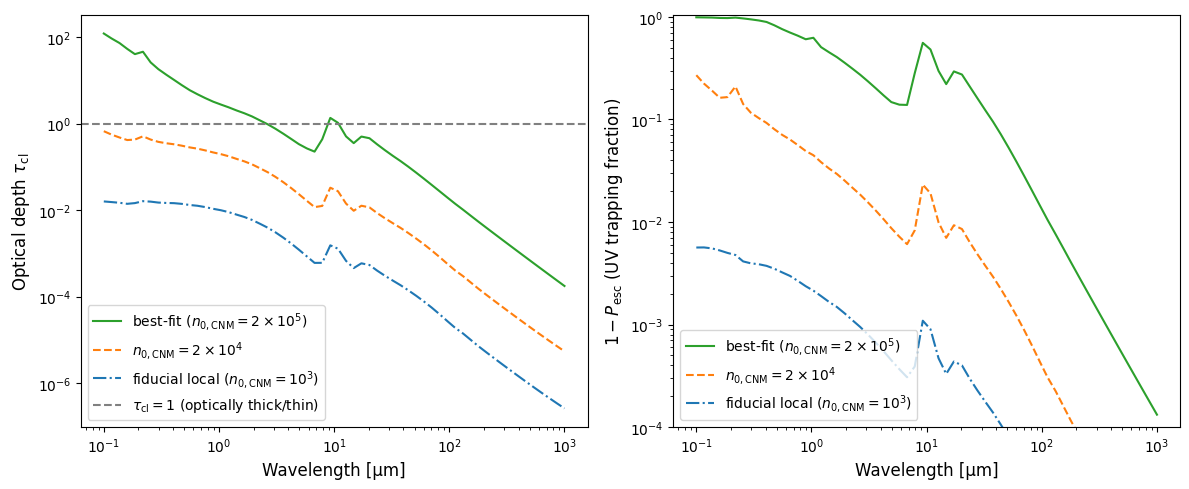}
    \caption{Wavelength dependence of the clump optical depth $\tau_{\rm cl}(\lambda)$ (left) and the
UV photon trapping fraction $1-P_{\rm esc}(\lambda)$ (right) at $t_{\rm gal}=80~\mathrm{Myr}$.
The solid green curve shows the best-fit model ($n_{0,\mathrm{CNM}}=2\times10^{5}~\mathrm{cm^{-3}}$),
the dashed orange curve shows an intermediate case ($n_{0,\mathrm{CNM}}=2\times10^{4}~\mathrm{cm^{-3}}$),
and the dot--dashed blue curve shows the fiducial local normalization
($n_{0,\mathrm{CNM}}=10^{3}~\mathrm{cm^{-3}}$), with all other parameters fixed to the best-fit values.
The horizontal grey dashed line in the left panel indicates $\tau_{\rm cl}=1$, separating optically
thick and thin regimes. Increasing $n_{0,\mathrm{CNM}}$ systematically increases  the optical depth in the UV and optical bands and enhances UV photon trapping, while the clumps remain optically thin in the FIR.
The feature around $\sim10~\mu$m reflects the silicate absorption band.}
    \label{fig:tau_pesc_n0}
\end{figure*}

Figure~\ref{fig:tau_pesc_n0} compares three cases at $t_{\rm gal}=80~\mathrm{Myr}$: the best-fit model
($n_{0,\mathrm{CNM}}=2\times10^{5}~\mathrm{cm^{-3}}$), an intermediate case
($n_{0,\mathrm{CNM}}=2\times10^{4}~\mathrm{cm^{-3}}$), and a fiducial ``local'' normalization
($n_{0,\mathrm{CNM}}=10^{3}~\mathrm{cm^{-3}}$). In this comparison, all other parameters are fixed to
the best-fit values, so that the differences directly isolate the impact of $n_{0,\mathrm{CNM}}$.
As $n_{0,\mathrm{CNM}}$ increases, $\tau_{\rm cl}(\lambda)$ systematically increases at UV/optical
wavelengths, reducing $P_{\rm esc}$ (increasing $1- P_{\rm esc}$) and thereby enhancing UV photon trapping. The best-fit case is
optically thick to stellar UV photons ($\tau_{\rm cl}(\lambda)\gtrsim 1$ in the UV), yielding
$P_{\rm esc}\ll 1$ and efficient UV photon trapping. In contrast, the fiducial local normalization
results in $\tau_{\rm cl}(\lambda)\ll 1$ over essentially all wavelengths and $P_{\rm esc}\approx 1$,
implying negligible trapping; the intermediate case falls between these two limits.

Importantly, even in the best-fit case the clumps become optically thin at far-infrared wavelengths
(e.g. $\lambda \gtrsim 50~\mu$m; $\tau_{\rm cl}\ll 1$), so the reprocessed FIR photons escape efficiently
($P_{\rm esc}\approx 1$). This confirms that the enhanced FIR output in our best-fit model is driven
by UV/optical photon trapping followed by IR re-radiation, rather than by FIR self-absorption.
The feature around $\sim 10~\mu$m (visible in both $\tau_{\rm cl}$ and $1- P_{\rm esc}$) reflects the
silicate absorption band in the adopted dust optical properties.

\bsp
\label{lastpage}
\end{document}